\documentclass[aps,prb,twocolumn,superscriptaddress,showpacs,showkeys]{revtex4-1}
\usepackage{graphicx}
\usepackage{amssymb}
\usepackage{booktabs}
\makeatletter
\def\@bibitemShut{}%
\mathchardef\@mmm=3000 %
\makeatother

\begin{document}
\title{Single-dopant resonance in a single-electron transistor}%
\author{V.~N. Golovach}%
\affiliation{SPSMS, UMR-E 9001, CEA-INAC/UJF-Grenoble 1, F-38054 Grenoble, France}%
\author{X. Jehl}%
\affiliation{SPSMS, UMR-E 9001, CEA-INAC/UJF-Grenoble 1, F-38054 Grenoble, France}%
\author{M. Houzet}
\affiliation{SPSMS, UMR-E 9001, CEA-INAC/UJF-Grenoble 1, F-38054 Grenoble, France}%
\author{M. Pierre}%
\affiliation{SPSMS, UMR-E 9001, CEA-INAC/UJF-Grenoble 1, F-38054 Grenoble, France}%
\author{B. Roche}%
\affiliation{SPSMS, UMR-E 9001, CEA-INAC/UJF-Grenoble 1, F-38054 Grenoble, France}%
\author{M. Sanquer}%
\affiliation{SPSMS, UMR-E 9001, CEA-INAC/UJF-Grenoble 1, F-38054 Grenoble, France}%
\author{L.~I. Glazman}%
\affiliation{Department of Physics, Yale University, New Haven, Connecticut 06520, USA}
\begin{abstract}%
Single dopants in semiconductor nanostructures have been studied in great details recently 
as they are good candidates for quantum bits, provided they are coupled to a detector. 
Here we report coupling of a single As donor atom to a single-electron transistor (SET) in a 
silicon nanowire field-effect transistor. Both capacitive and tunnel coupling are achieved, 
the latter resulting in a dramatic increase of the conductance through the SET, by up to one order of magnitude. 
The experimental results are well explained by the rate equations theory developed in parallel with the experiment.
\end{abstract}
\maketitle

\section{Introduction}
\label{secIntro}

Metallic single-electron transistors (SETs) have found several applications 
including charge detectors and amplifiers~\cite{Schoelkopf98,Devoret00},
electron pumps for metrology~\cite{Pothier92,Keller99},
and low-temperature thermometers~\cite{Pekola94}.
More recently, single dopants in silicon have emerged as promising candidates for qubits of the ultimately small size~\cite{Kane98}.
Silicon technology offers the opportunity both to realize classical SETs~\cite{ScottThomas,Takahashi,Ali,Leobandung,Fujiwara,Hofheinz06,Angus} 
and to probe single dopants~\cite{Sellier,Ono,Calvet,Pierre10,Tan,Morello}.
The use of the SET for quantum-mechanical measurement of the dopant qubit 
requires characterization of the coupling strength between SET and dopant atom.
Here we study the mutual capacitive and tunnel couplings between a dopant atom and a SET, 
realized in a silicon nanowire field-effect transistor.

The interplay between Coulomb interaction and tunneling is most thoroughly studied in mesoscopic physics~\cite{Aleiner02},
where it leads to the so-called Coulomb blockade (CB) oscillations~\cite{KulikShekhter,FultonDolan} 
of conductance, commonly observed in metallic SETs~\cite{KastnerReviewAnPh} and semiconductor quantum dots~\cite{KastnerReviewAnPh,KAT}.
The energy scale governing the CB oscillations in a metallic SET is set by the charging energy 
$E_C=e^2/2C$, where $C$ is the total capacitance of the SET island relative to the surrounding conductors, 
and $-e$ is the electron charge.
At temperatures $T\sim E_C$, the conductance as a function of the gate voltage acquires an oscillatory component
and, with lowering the temperature, it turns into a series of spikes (CB peaks), at $T\ll E_C$.
At the lowest temperatures,
the height of a CB peak depends on the properties of the electron wave function inside the SET; 
namely, on its values at the contacts.
The peak height changes randomly from peak to peak, due to the chaotic nature of the electron motion (and wave functions) inside the SET.
Such mesoscopic fluctuations of the CB peaks persist up to temperatures on the order of the single-particle level spacing, $\Delta E$, of the SET.
It is convenient to distinguish between SETs and quantum dots by the magnitude of $\Delta E$:
SETs are devices in which $\Delta E\ll E_C$ and an intermediate temperature regime, $\Delta E\ll T\ll E_C$, is possible, 
whereas quantum dots are devices in which $\Delta E\sim E_C$.
The existence of the temperature regime $\Delta E\ll T\ll E_C$ makes SETs outstanding candidates for applications.

In the temperature regime $\Delta E\ll T\ll E_C$, the mesoscopic fluctuations of the CB peaks are suppressed.
The CB peaks have equal height, which is determined by the transparencies of the SET tunnel barriers,
and thus, it is independent on the details of the electron motion inside the SET island.
The peak height remains constant over a large number of CB oscillations and
may vary weakly with the gate voltage due to the way the gate couples to the SET tunnel barriers.
A strong modulation of the peak height on a scale of several CB oscillations is usually considered as
an anomaly in this regime.
Such anomalies in the CB oscillations have been attributed to charge traps,
which may be present in the SET surrounding~\cite{Grupp2001,HofheinzEPJ}.
Charge traps is a common and, typically, unwanted ingredient in modern electronic nanodevices.
Their most conspicuous collective effect is the voltage switch, which makes devices difficult to control.  
In silicon nanostructures based on doping modulation, charge traps occur naturally at the borders of the doped regions, 
where the dopants are likely to diffuse away from their majority distribution and form nearly isolated charge traps.
In this paper, we study the effects associated with a single charge trap (a dopant atom) in the CB oscillations of a SET.

We observe an interplay between mutual capacitive coupling and tunneling 
in the silicon SET coupled to a dopant atom.
In particular, we observe a spectacular enhancement of the conductance through the SET
when transport occurs by resonant tunneling via the dopant atom.
We develop a theory which takes into account both tunneling and capacitive coupling
between dopant atom and SET, and assess the coupling strengths by analyzing 
the CB oscillations of the linear conductance.
The mutual capacitive coupling between SET and dopant atom, extracted from the experiment, 
has an energy scale of $\sim 1\,\textrm{meV}$, which indicates a good charge sensitivity of the SET.
In this double-dot system, i.e. the system of coupled dopant atom and SET, 
the SET can be used for manipulation and readout of the single dopant atom.

The paper is organized as follows.
In Sec.~\ref{secMainRes}, we present the experimental results alongside with a qualitative explanation of the observed effects
and a quantitative fit to the experimental data using the theory developed in subsequent sections.
We also formulate a simple electrostatic model which helps in discussing the experimental findings.
In Sec.~\ref{secDerivMnRs}, we proceed with a systematic analysis of the electrostatic model.
We first consider the low-temperature limit, see Sec.~\ref{secResDerElMod},
in which transport is activational, and hence, governed by the energy spectrum derived from the electrostatic model.
Then, we complete the electrostatic model by kinetic-energy and tunneling terms and 
proceed with the study transport through the coupled donor-SET system by means of rate equations, see Sec.~\ref{secTransport}.
We find that at low temperatures, the linear conductance could also be derived from a circuit approach,
in which the donor-SET system is replaced by a circuit of resistors.
We analyze the validity of this approximation at higher temperature by solving the rate equations numerically
and find good agreement for the parameter sets which we extracted from fitting to the experimental data.
In Appendix~\ref{appStabilityDiagram}, we give the necessary formulae for a quantitative description of 
the double-dot stability diagram occurring in the electrostatic model.

\section{Main Results}
\label{secMainRes}

\begin{figure*}[th!]
\begin{center}
\begin{minipage}[t]{0.530101\textwidth}
\vspace{0pt}\includegraphics[width=1.0\textwidth]{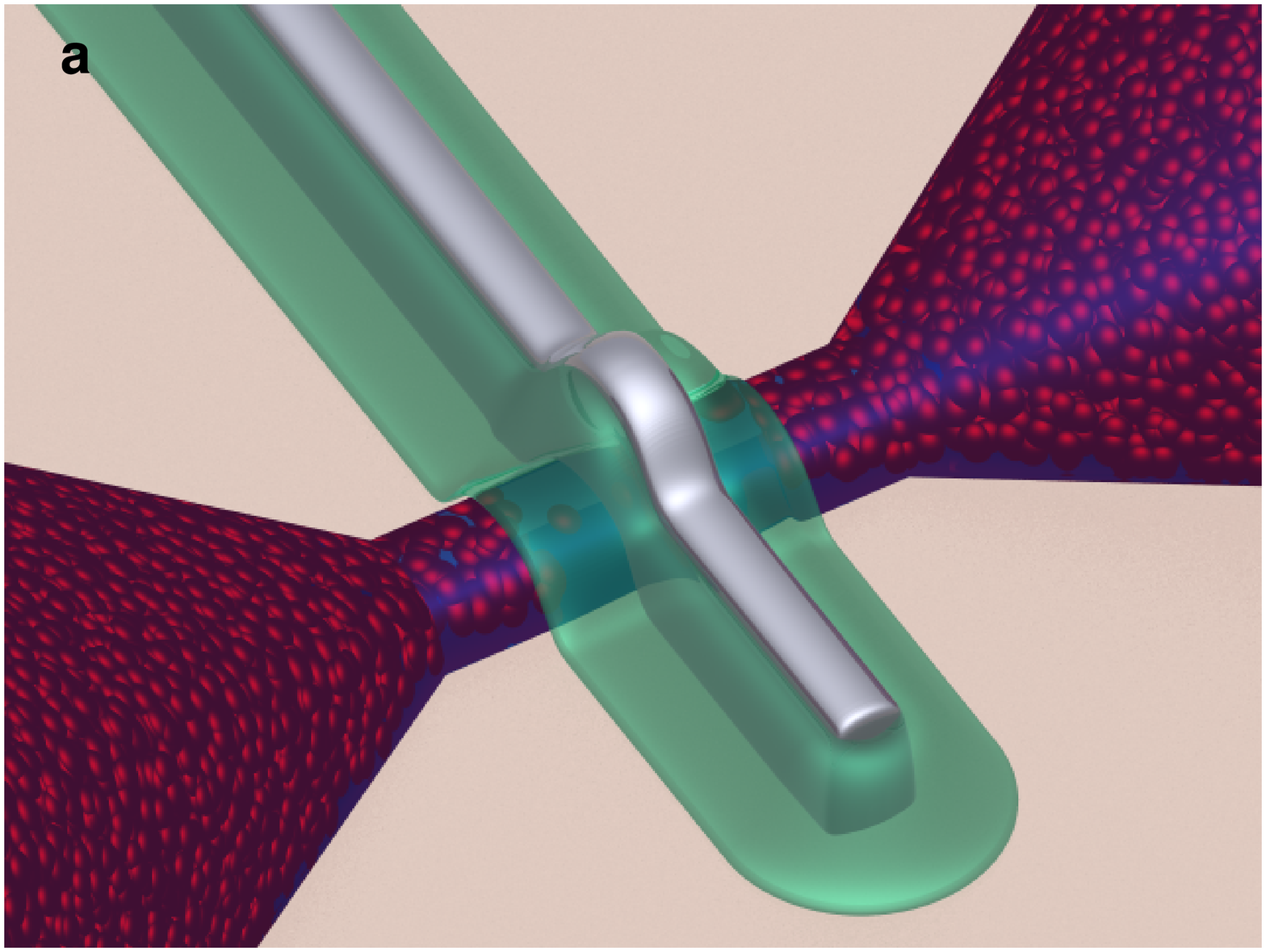}
\end{minipage}
\hspace{5pt}
\begin{minipage}[t]{0.459899\textwidth}
\vspace{0pt}\includegraphics[width=1.0\textwidth]{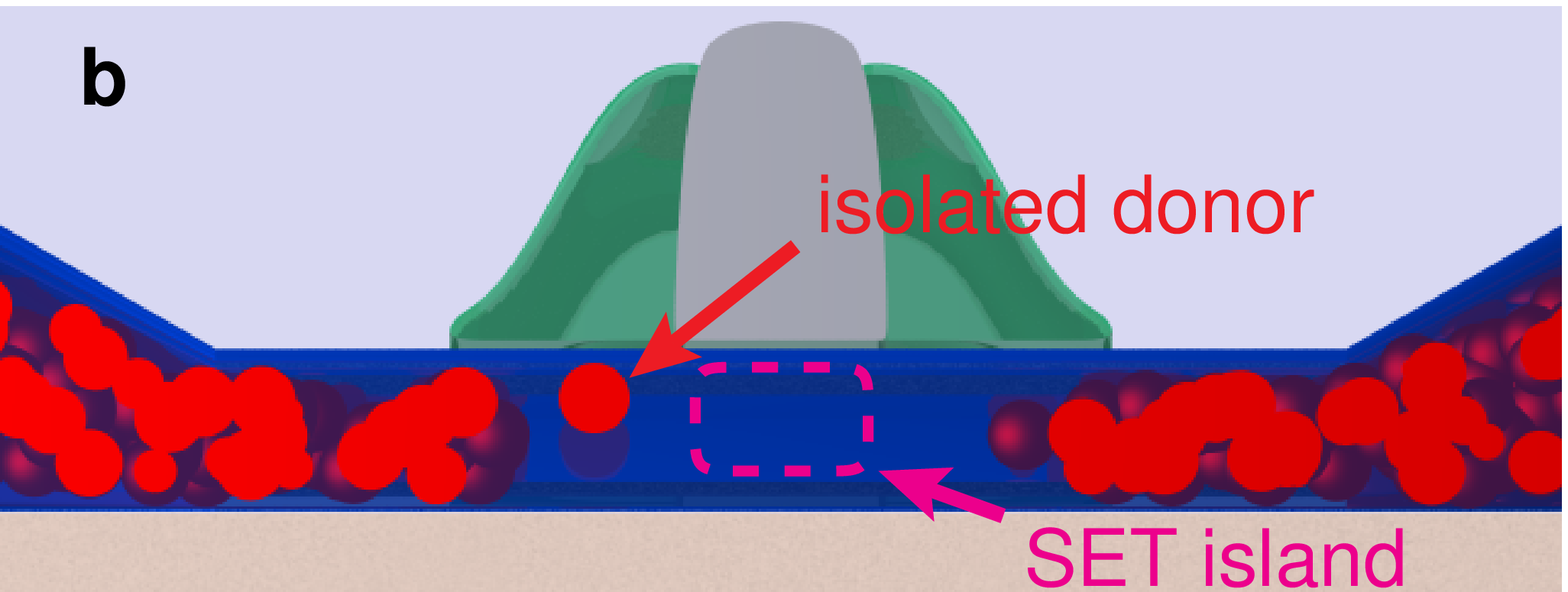}\par
\vspace{2pt}\includegraphics[width=0.420655\textwidth]{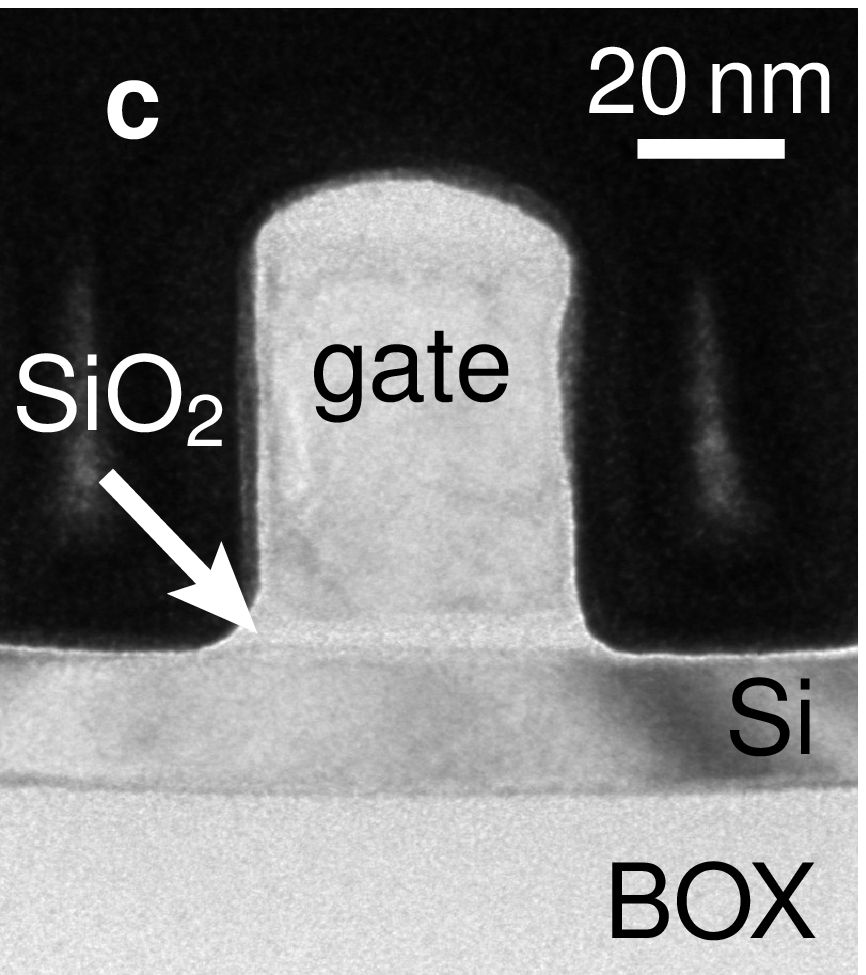}
\hfill
\vspace{0pt}\includegraphics[width=0.559345\textwidth]{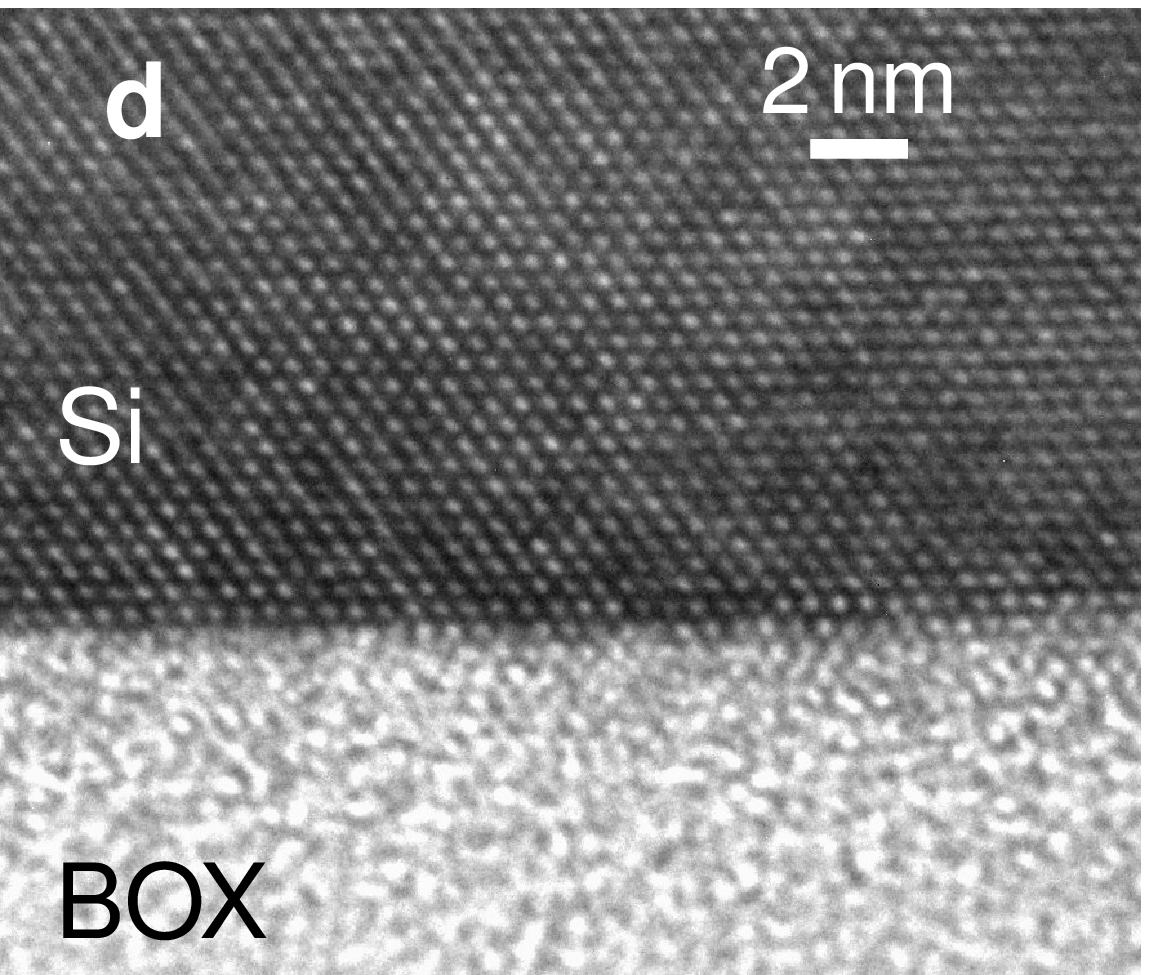}
\end{minipage}
\caption{
Gated nanowire device.
(a)
Geometrical arrangement of the main components of the device.
The Si nanowire channel ($20\,\textrm{nm}$ thick and $60\,\textrm{nm}$ wide) 
bridges the constriction between the source and drain.
The gate (40-nm-wide poly-silicon finger)
is deposited on top of the nanowire and it is
surrounded by silicon nitride ($\textrm{Si}_3\textrm{N}_4$), 
thus forming two spacers (each $40\,\textrm{nm}$ wide) on both sides of the gate.
The red spheres illustrate the doping with As atoms inside the Si material; 
dopants on the surface of Si are omitted for clarity.
(b)
Cross section along the nanowire.
In the experiment, the implantation with As atoms is made after the deposition of gate and spacers.
Therefore, during the implantation, a small number of As atoms may diffuse into the region under the spacer, forming isolated donors.
The SET island is formed by electron accumulation when a positive voltage is applied to the gate.
(c)
Transmission-electron-microscopy image of a cross section along the nanowire in an actual device.
The nanowire (Si) is separated from the poly-silicon finger (gate) by
a $4\,\textrm{nm}$ thick oxide layer ($\textrm{SiO}_2$); the spacers are hard to discern on the micrograph.
(d)
High-resolution image showing the atomic lattice of the nanowire and its interface with the buried oxide (BOX).
}
\label{figsamples}
\end{center}
\end{figure*}

We implement an electrostatically-defined SET in a Si nanowire fabricated using 
the industrial etching and doping techniques described in Ref.~\onlinecite{Hofheinz06}.
A schematic view of the device is shown in Fig.~\ref{figsamples}a.
The SET island is formed by electron accumulation under a top gate 
in a region of intrinsic Si, see Fig.~\ref{figsamples}b.
The leads are defined by doping with As atoms (donors). 
40-nm-wide spacers, made of silicon nitride, 
are deposited on both sides of the gate prior to the ion implantation 
in order to create low-doped regions acting as tunnel barriers between the leads and the island, see Figs.~\ref{figsamples}a and~\ref{figsamples}b.
Indeed, samples without spacers do not exhibit SET behavior, but allow to study transport 
through single dopants which diffused away from the leads 
into the Si region below the gate~\cite{Pierre10}.
A large number of devices with spacers ($\sim 10^4$) have been fabricated 
and more than $50$ have been studied at low temperatures, $T\geq 40\,\textrm{mK}$. 
All devices exhibit low $1/f$ noise and very periodic CB oscillations. 

\begin{figure*}[th!]
\begin{center}
\includegraphics[width=0.69\textwidth]{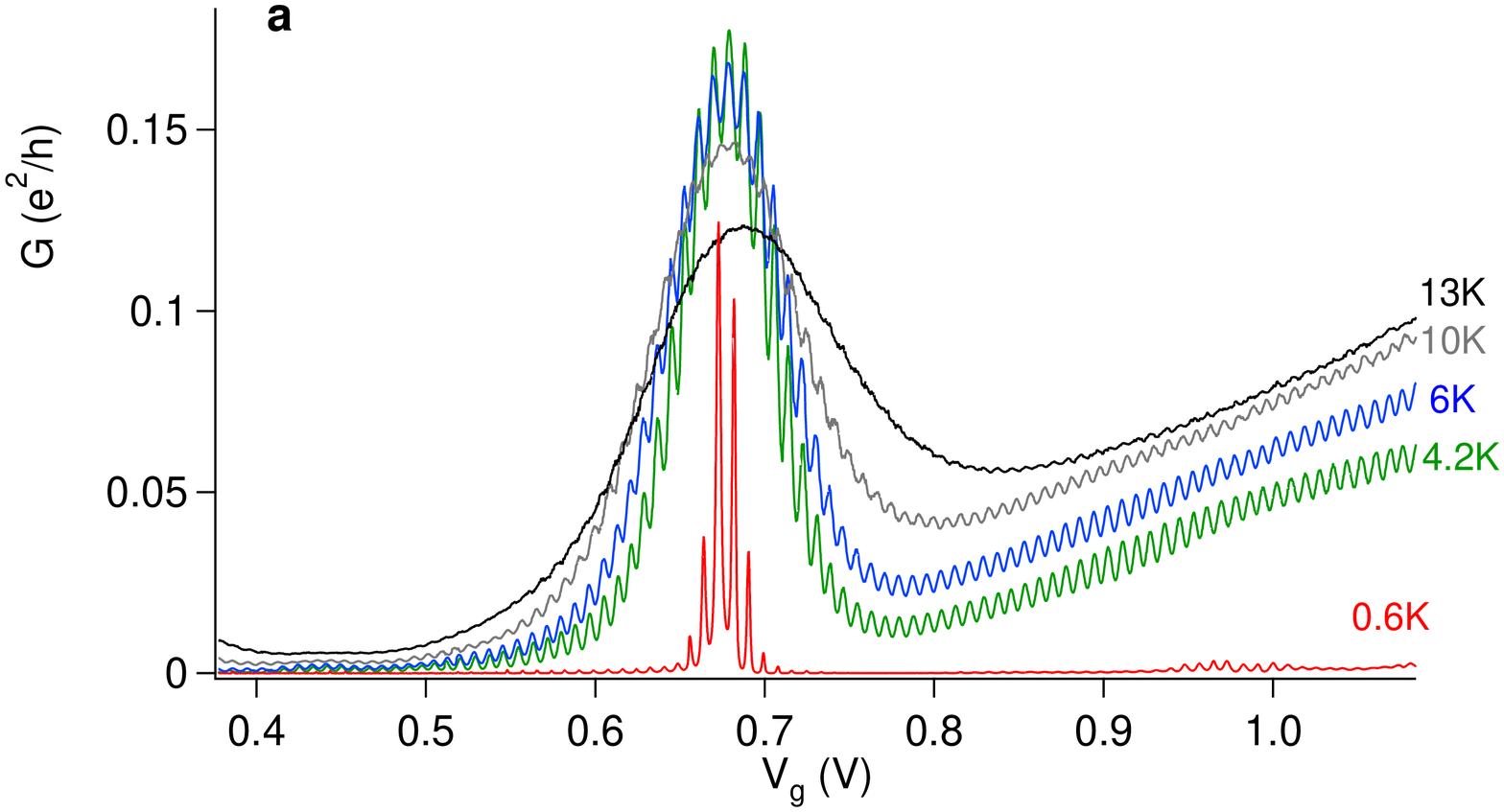}
\includegraphics[width=0.3\textwidth]{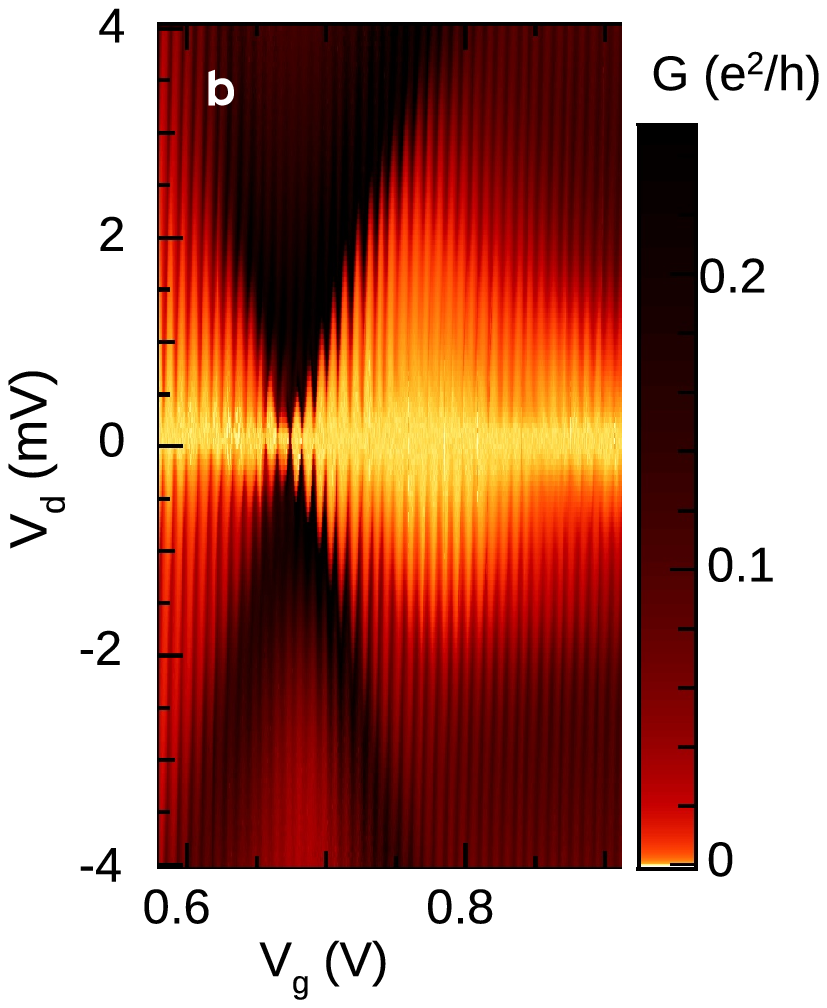}
\caption{
Experimental signatures of tunnel coupling between a single dopant atom and a single-electron transistor.
(a)
Linear conductance through the silicon nanowire as a function of the gate voltage $V_g$ at various temperatures.
The strong, resonance-like modulation of the CB oscillations is attributed to the presence of a dopant atom in the
SET barrier.
(b)
Color plot of the differential conductance of the device versus the gate and bias voltages.
The small Coulomb diamonds of the SET are strongly modulated in intensity,
revealing, on a larger scale, resonant tunneling associated with the dopant atom.
}
\label{figdata}
\end{center}
\end{figure*}

The high quality of the SET, which can be inferred from Figs.~\ref{figsamples}c and~\ref{figsamples}d, allows us to focus on special regions of gate voltage, 
in the trace of CB oscillations, where the transport properties of the SET are strongly perturbed by single As atoms.
We observe between $1$ and $5$ of such {\em anomalies} in each device in different ranges of the top-gate voltage $V_g$.
In previous reports of anomalous behaviour of the SET~\cite{HofheinzEPJ,Pierre09,George10,Huebl10},
the coupling between the SET island and the single dopants has been predominantly capacitive.
Here we observe dramatic increase of the current through the SET due to resonant tunneling through single dopants.
In Fig.~\ref{figdata}, we show the signatures of tunnel coupling of an As atom to the lead and SET island;
resonant features are observed both in the linear and differential conductances.

\begin{figure*}[th!]
\begin{center}
\includegraphics[width=0.49\textwidth]{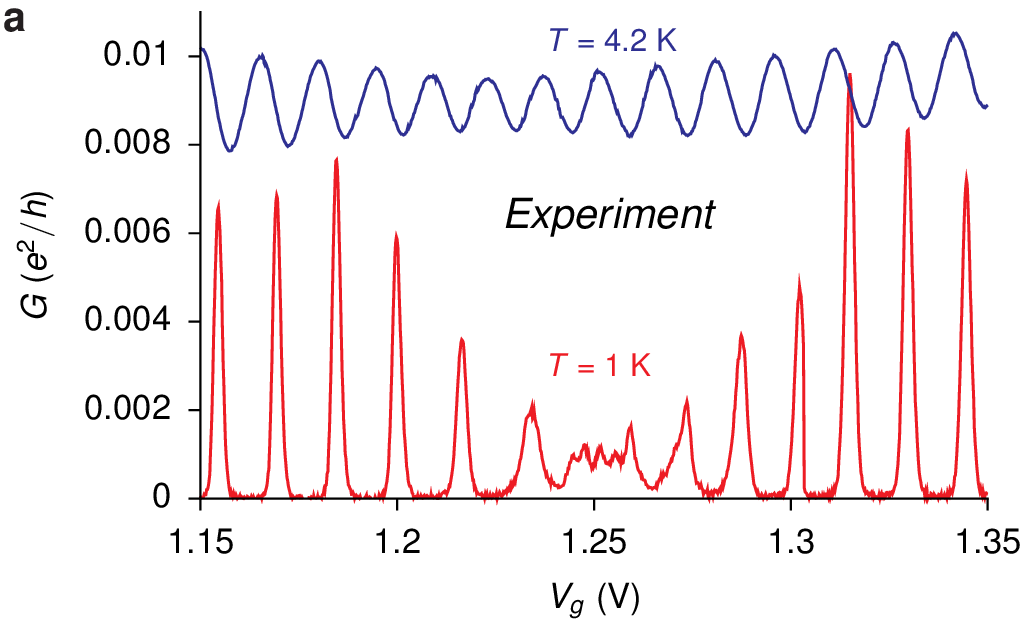}
\includegraphics[width=0.48\textwidth]{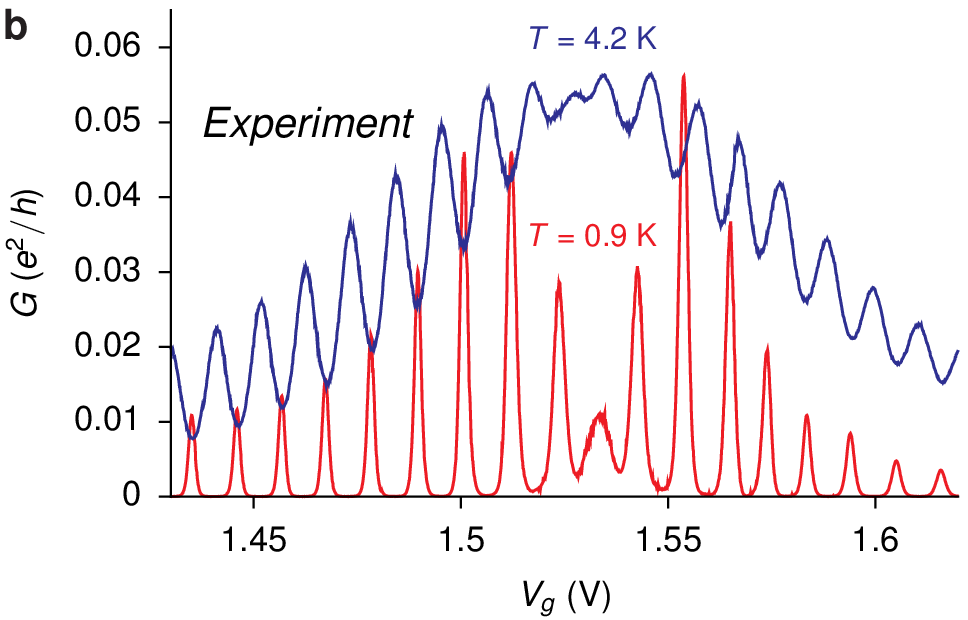}
\\
\includegraphics[width=0.49\textwidth]{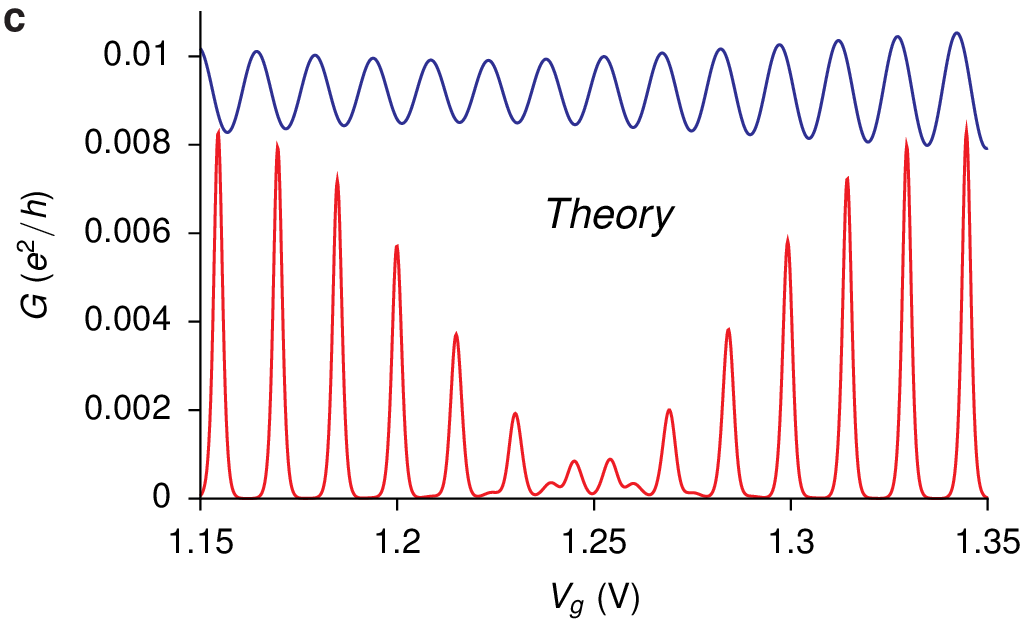}
\includegraphics[width=0.48\textwidth]{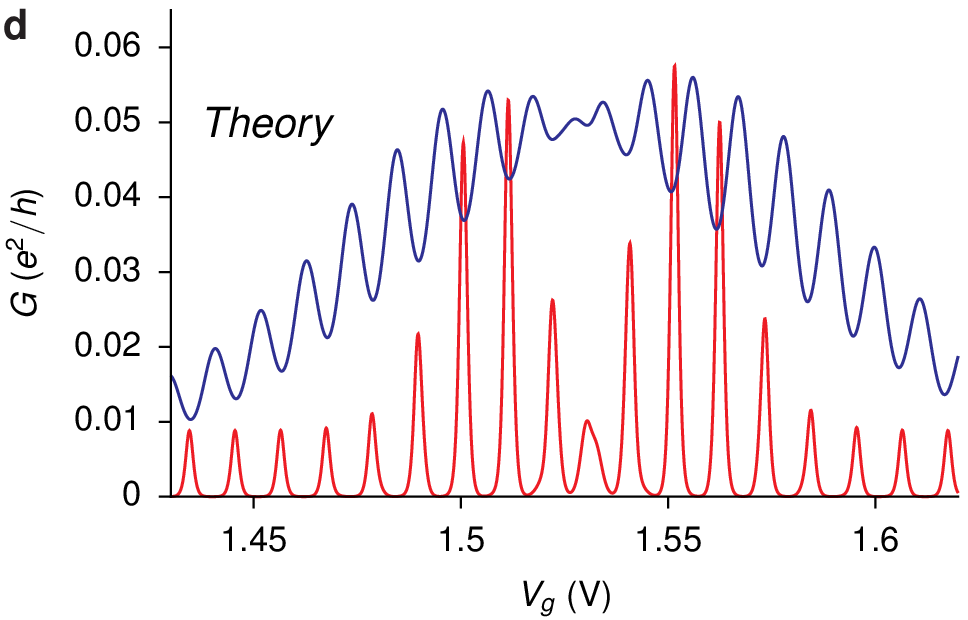}
\caption{
Anomalies in transport through SETs caused by single dopant atoms.
(a)
Linear conductance measured in a sample in which the dopant atom couples mostly capacitively to the SET.
At $T=1\,\textrm{K}$,
the CB peaks are suppressed in an interval of $V_g$, marking the extension of the anomaly.
At $T=4.2\,\textrm{K}$,
the CB oscillations are weakly modulated.
(b)
Modulation of CB oscillations in a different device, in which
the dopant atom is tunnel-coupled to both the SET island and the lead.
The envelope of CB oscillations reveals resonant tunneling through the dopant atom.
At $T=0.9\,\textrm{K}$, two resonances, separated by a suppression region, are visible in the envelope function.
At $T=4.2\,\textrm{K}$, 
the resonances overlap and form a single broad resonance due to tunneling via the dopant atom.
Note the factor-of-$5$ difference in the scale on the ordinate axis as compared to (a).
(c)
Linear conductance calculated using a circuit approach, see Sec.~\ref{secSumOfResFrom}.
With the donor coupling parameters $U_{12}=1.2\,\textrm{meV}$ and  $\Gamma_L=\Gamma_R=0$,
the experiment in (a) is reproduced almost quantitatively.
(d) 
The same as in (c), but plotted using the coupling parameters
$U_{12}=1.15\,\textrm{meV}$ and 
$\Gamma_L=\Gamma_R=40\,\mu\textrm{eV}$,
which nears the result to (b).
The SET parameters used in (c) and (d) are tabulated in Table~\ref{Table1}.
}
\label{fig1}
\end{center}
\end{figure*}

We focus on the linear conductance and analyze two representative types of anomalous behaviour
observed in the experiment, see Figs.~\ref{fig1}a and~\ref{fig1}b.
The anomalous behaviour has the following generic pattern:
(i) at the lowest temperatures, the CB peaks are strongly suppressed within
an interval of $V_g$,
(ii) at intermediate temperatures,
two maxima develop in the envelope of the CB oscillations
on both sides of the interval of suppression,
(iii) at higher temperatures,
the region of suppressed conductance turns into a region of
elevated conductance, with a single maximum in the envelope.
This pattern, evidenced by Fig.~\ref{fig1}b, is generic to the case when both capacitive and tunnel couplings are present.
The measurements shown in Fig.~\ref{fig1}a are consistent with having only capacitive coupling.
In this case, the envelope of CB oscillations exhibits no maxima; 
the anomalous behaviour consists in (i) only.

We successfully reproduce the anomalous behaviour of the CB oscillations
observed in the experiment, taking into account capacitive and tunnel couplings
of the dopant atom to the SET.
The results of our theory are shown in Figs.~\ref{fig1}c and~\ref{fig1}d,
and correspond, respectively, to Figs.~\ref{fig1}a and~\ref{fig1}b of the experiment.
In the following paragraphs, we formulate a simple model for the electrostatic energy
of the device and explain the observed behavior in terms of
sequential tunneling of electrons between energetically favorable states.

\begin{figure}[th!]
\begin{center}
\includegraphics[width=0.4\textwidth]{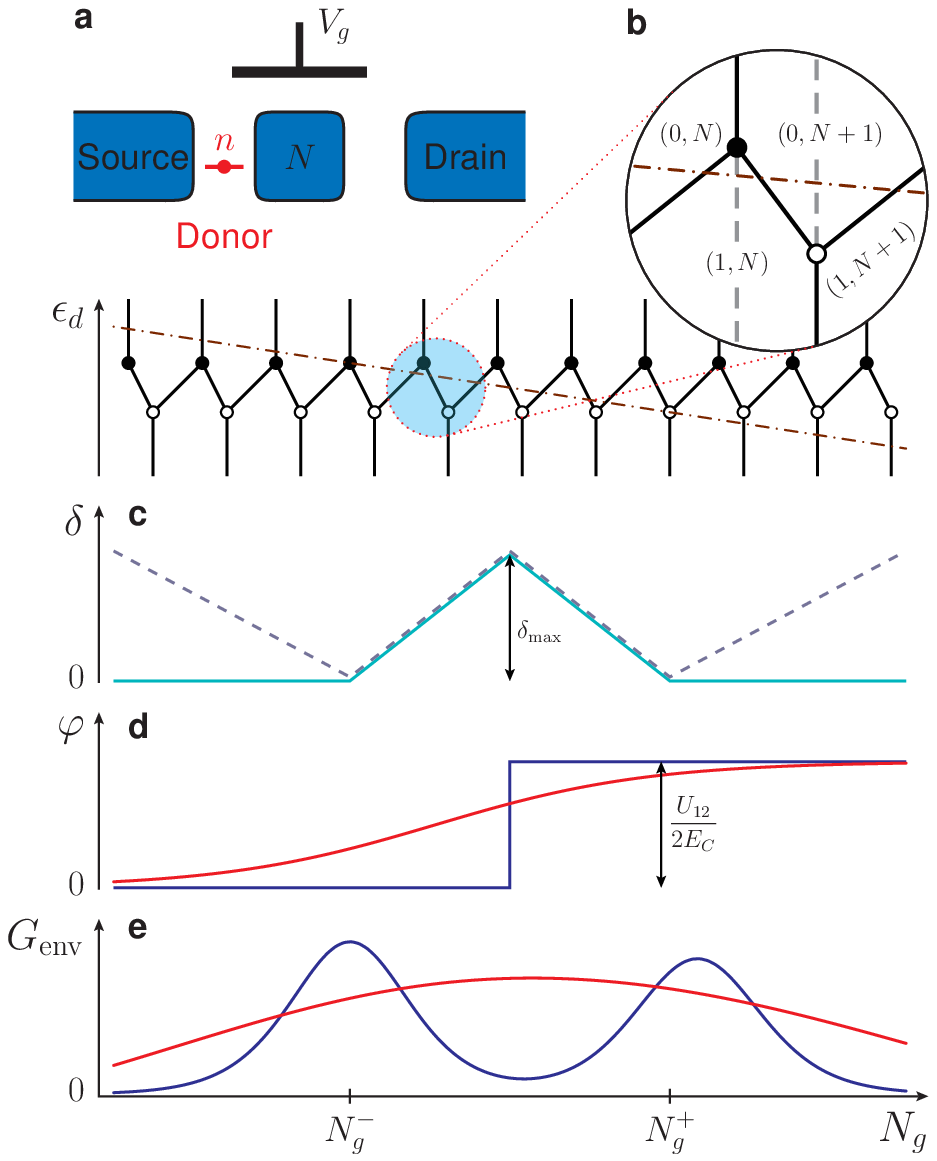}
\caption{
Setup, stability diagram, energy gap, phase shift, and envelope of CB oscillations.
(a)
Schematic view on the device. 
The SET island, with charge $N$, is tunnel-coupled to two leads (source and drain).
The donor, with charge $n$, is situated in the tunnel barrier between the source and the SET island.
The top gate, at voltage $V_g$, couples capacitively to the SET island and donor; 
the coupling to the latter is the weakest of the two.
(b)
Stability diagram of the donor-SET system.
Solid lines separate regions with different ground state charge configurations.
Sweeping the gate voltage $V_g$ in the experiment corresponds to traversing the 
plane of the stability diagram along the dash-dotted line -- the ``working line'' of the device.
The anomaly occurs in the region between the upper and lower rows of triple points. 
For small coupling of the donor to the gate ($\alpha\ll 1$), 
the working line is nearly horizontal; the anomaly region comprises, then, a large number of 
charge configurations. 
(c)
The energy gap, governing the activated conductance at CB peaks, as a function of the gate voltage $N_g$.
The solid line corresponds to the case $g_L\neq 0$, whereas the dashed line to $g_L=0$.
(d)
The phase shift $\varphi$, which gives the position of the CB peaks across the anomaly region and outside.
At low temperatures, $\varphi$ changes stepwise in the center of the anomaly.
At high temperatures, $\varphi$ changes gradually over an interval which grows proportionally to the temperature.
(e)
The envelope of the CB oscillations versus the gate voltage.
At low temperatures, two resonances occur on the sides of the anomaly.
At high temperatures, the resonances overlap, forming a single broad resonance.
}
\label{fig2}
\end{center}
\end{figure}

The dopant atom (donor) resides in one of the SET barriers;
let it be the barrier next to the source.
A sketch of the setup is shown in Fig.~\ref{fig2}a.
The energy associated with the charges in the device
depends on the donor charge $n=0,1$ and the SET charge $N$
(both measured in units of the electron charge, $-e$), and
can be abbreviated in the following form
\begin{equation}
E(n,N)=\epsilon_d n + E_C(N-N_g)^2 + U_{12}n(N-N_g),
\label{eqelectrostatic}
\end{equation}
where $\epsilon_d$ is the energy of the donor level,
$E_C$ is the charging energy of the SET,
$N_g=C_gV_g/e$ is the dimensionless gate voltage,
with $C_g$ being the gate-to-SET capacitance,
and $U_{12}$ is the energy of mutual capacitive coupling.
The top gate couples to both the SET and the donor.
However, the lever arm of the donor is considerably smaller
than that of the SET, because 
the donor is more distant from the gate than the SET island is,
and the donor charge is partly screened due to proximity to the lead.
We express this fact in the relation $\epsilon_d= -\alpha E_C N_g+\mbox{const}$, with
$\alpha\ll 1$.

At low temperatures, transport occurs due to 
transitions from the ground state
to the lowest-in-energy charge configurations.
The stability diagram, commonly used for double dots~\cite{VanderWiel},
shows the ground state as a function of $\epsilon_d$ and $N_g$, see Fig.~\ref{fig2}b
and Appendix~\ref{appStabilityDiagram}.
Variation of the gate voltage $V_g$ leads to traversing the plane
of the stability diagram along an inclined line --- the ``working line'' of the device.
In Fig.~\ref{fig2}b, the dash-dotted line represents the working line.
Each time the working line crosses a solid line in the stability diagram,
two charge configurations have equal energies in Eq.~(\ref{eqelectrostatic}).
(For simplicity of argument, we measure single-particle energies from the Fermi level in the leads.)
The conductance, however, may display CB peaks only for certain types of degeneracies.

We consider first the effect of direct tunneling through the SET,
which does not involve a change of the donor charge.
The activation energy of this process is determined by the difference
$E(n,N)-E(n,N+1)$. 
Note that both terms in the latter expression are evaluated at 
the same $n$, which corresponds to the condition that there is no current
through the donor.
That energy difference reaches zero every time the
working line of the device intersects a vertical line of the
stability diagram, which is possible for $N_g<N_g^-$ and $N_g>N_g^+$, see Fig.~\ref{fig2}b.
These intersections correspond to CB peaks which have
temperature-independent heights (zero activation energy, see Fig.~\ref{fig2}c). 
For $N_g<N_g^-$, the conductance around the peaks can be approximated by~\cite{GlazmanShekhter}
\begin{equation}
G(N_g)\simeq \sum_N{\cal G}(x-x_N),\quad
{\cal G}(z)=\frac{e^2}{h}\frac{g_Lg_R}{g_L+g_R}\frac{z}{\sinh(z)},
\label{eqlinearG}
\end{equation}
with $x=2E_C N_g/T$ and $x_N=2E_C (N+1/2)/T$.
Here,
the summation goes over integer $N$,
the function ${\cal G}(z)$ describes the shape of a single peak,
$g_L$ and $g_R$ are the dimensionless conductances,
in units of $2e^2/h$, of the SET tunnel barriers~\cite{Aleiner02}.
Equation~(\ref{eqlinearG}) describes transport due to sequential tunneling at
low temperatures ($T\ll E_C$) and for positions in the stability diagram 
far above the triple points, where the donor degree of freedom is frozen to $n=0$.
For $N_g>N_g^+$, far below the triple points,
where the donor degree of freedom is frozen to $n=1$,
the conductance is given by Eq.~(\ref{eqlinearG}), with $x_N\to x_N+U_{12}/T$.
The periodic CB oscillations on both sides of the
anomaly are shifted with respect to each other by a fraction, $U_{12}/2E_C$,
of the period.
Using this fact, we extract 
$U_{12}= 1.2\pm 0.1\,\textrm{meV}$ 
for the sample in Fig.~\ref{fig1}a and 
$U_{12}= 1.15\pm 0.1\,\textrm{meV}$
for the sample in Fig.~\ref{fig1}b.

Charge-charge correlations are essential to understanding the transport
mechanism in the anomaly region, $N_g^-<N_g<N_g^+$.
The donor charge adjusts to the SET charge, lowering
the electrostatic energy in Eq.~(\ref{eqelectrostatic}).
This effect is strongly pronounced around the SET resonances,
at which the difference $E(n,N)-E(n,N+1)$ vanishes, see dashed lines in inset of Fig.~\ref{fig2}b.
For concreteness, let us discuss the dashed lines of the upper triple points, 
where $E(0,N)=E(0,N+1)$.
The lowest-in-energy charge configuration at these points is $(1,N)$, whereas the
configurations $(0,N)$ and $(0,N+1)$ correspond to excited states.
In order for the transport to occur, the donor charge has to change from
$n=1$ to $n=0$.
This process is suppressed by an activation energy
$\delta=E(0,N)-E(1,N)$, evaluated at the SET resonance $E(0,N)=E(0,N+1)$.
The presence of a finite activation energy is responsible for the behavior of type
(i) observed in the experiment.
The height of a CB peak in the anomaly region decreases with lowering the
temperature as $\propto\exp(-\delta/T)$, where $\delta$ is shown in Fig.~\ref{fig2}c
for the whole range of the gate voltage.
Note that the gap $\delta$ reaches maximum, 
$\delta_{\rm max}=(U_{12}/2)(1-U_{12}/2E_C)$, 
in the middle of the anomaly
and it goes to zero towards the edges.
We analyze the conductance around a CB peak in Sec.~\ref{secDerivMnRs}.
At low temperatures ($T\ll E_C$),
the conductance can be written as
\begin{eqnarray}
G(N_g)&=& \sum_N
\left[
(1-\langle n\rangle)
{\cal G}(x-x_N)\right.\nonumber\\
&&\left.
+
\langle n\rangle
{\cal G}(x-x_N-U_{12}/T)
\right],
\label{linGnaver}
\end{eqnarray}
where $\langle n\rangle$ is the average occupation of the donor.
The occupation $\langle n\rangle$ 
varies strongly with the gate voltage inside the anomaly region,
whereas it approaches a constant value of $0$ or $1$ outside.
We give explicit expressions for $\langle n\rangle$ in the vicinity of triple points in Sec.~\ref{supplSecnaver}.
Equation~(\ref{linGnaver}) has a simple physical meaning: 
the donor charge shifts the gate voltage ``seen'' by the electron passing through the SET; 
the probabilities of having the donor charge $1$ and $0$ are, respectively,
$\langle n\rangle$ and $1-\langle n\rangle$.

The CB peaks in the anomaly region
described by Eq.~(\ref{linGnaver})
occur at shifted positions, $N_g=N+1/2+\varphi$, where
$\varphi$ is a phase shift of CB oscillations.
We show the behaviour of $\varphi$ at low and high temperatures in Fig.~\ref{fig2}d.
At $T\gg U_{12}$, the phase shift is given by
$\varphi=\langle n\rangle U_{12}/2E_C$, where
$\langle n\rangle$ assumes the form of the average occupation of an isolated donor.
In addition to the CB peaks being shifted in the anomaly region,
they also become broader than usual.
Usually, the width of a CB peak is given by $\Delta N_g \propto T/E_C$,
which is a relation commonly used in determining the electron temperature in the sample.
We find that, at $T\lesssim\delta$, this relation is no longer valid
and one has $\Delta N_g \propto \min(T/U_{12},\delta/E_C)$.
With the peaks being broader in the anomaly region, the conductance in the 
middle of CB valleys becomes larger.
Indeed, instead of the usual suppression of the conductance in the middle of the valley, $\propto\exp(-E_C/T)$,
we find a smaller suppression factor, $\propto\exp(-\overline{\delta}/T)$, with the activation energy 
$\overline{\delta}\leq E_C$ 
reaching its minimum, $\overline{\delta}_{\rm min}=E_C-U_{12}/2$, in the middle of the anomaly.

We concentrate now on the role of tunneling through the donor and, for
that purpose, we dispense with the possibility of direct tunneling between the 
SET and the source ($g_L=0$).
To have an activationless transport in that case one has to satisfy simultaneously two
conditions for the energy: $E(n,N)=E(n,N+1)$ and $E(0,N+1)=E(1,N)$.
Clearly these two conditions may be satisfied only at the triple points. 
We consider a small $\alpha$ for which the working line passes 
by triple points sufficiently close ($\alpha\ll T/E_C$).
The maxima of the CB peaks form a smooth envelope function, $G_{\rm env}(N_g)$,
which contains information about the tunnel coupling of the donor.
Indeed, we find that, in the vicinity of the upper row of triple points,
\begin{equation}
G_{\mathrm{env}}(N_g)\simeq \frac{e^2}{h}\frac{\pi\Gamma_L\Gamma_R}{T(\Gamma_L+\Gamma_R)}
\frac{1}{\cosh^2(y/2)},
\label{eqrestunY}
\end{equation}
with $y=\alpha E_C(N_g-N_g^-)/T$ and 
$\Gamma_L$ ($\Gamma_R$) being the tunneling rate between donor and source (SET island).
Equation~(\ref{eqrestunY}) coincides with the conductance through a resonant level.
In deriving Eq.~(\ref{eqrestunY}), we made several simplifying assumptions (see Sec.~\ref{SupplSec1TrPt}),
which do not change the essence of our result:
the information about the donor is encoded in the envelope of the CB oscillations.
At low temperatures, $T\ll U_{12}$, the envelope function consists of two resonances 
centered at $N_g=N_g^\pm$, see Fig.~\ref{fig2}e.
The resonance at $N_g=N_g^+$ differs slightly from the one at $N_g=N_g^-$, because of
the spin degeneracy on the donor.
Each of the resonances corresponds to traversing a horizontal
line of triple points in Fig.~\ref{fig2}b.
The behavior of type (ii) observed in the experiment (see Fig.~\ref{fig1}b) 
is, thus, due to resonant tunneling through the donor level.
At high temperatures, $T\gg U_{12}$, the two resonances overlap and 
merge into a single broad feature, see Fig.~\ref{fig2}b,
corresponding to the behavior of type (iii) seen, e.g., in Fig.~\ref{fig1}b.

Thus far, we have separated the effect of tunneling through the donor from
direct tunneling by setting $g_L=0$.
In this special case, the CB peaks are suppressed at low temperatures 
both inside and outside the anomaly region.
The corresponding activation energy $\delta$ is shown by the dashed line in Fig.~\ref{fig2}c.
At a finite but small $g_L$, the envelope maxima around $N_g=N_g^\pm$ remain present, but
the smaller activationless CB peaks are visible outside the interval $N_g^-<N_g<N_g^+$.
The activation energy $\delta$ is, then, given by the solid line in Fig.~\ref{fig2}c.
In order to describe the interplay between direct tunneling and tunneling via the donor,
we employ a circuit approach~\cite{Ruzin}.
We associate a resistor with each tunnel junction and
sum up the resistances of the so-obtained circuit 
according to the classical laws for circuits, see Sec.~\ref{secSumOfResFrom}.
The end result is a conductance formula, which we use to reproduce the
experimental data in Fig.~\ref{fig1}.
We also back up the circuit approach by an exact numerical evaluation of the conductance.
We find that the circuit approach works extremely well for the experimentally relevant parameters.

\section{Derivation of Main Results}
\label{secDerivMnRs}
In this section, we consider the transport through the donor-SET system in greater detail 
and derive the results presented in Sec.~\ref{secMainRes} regarding our theoretical description of transport.

\subsection{Activation energies derived from the electrostatic model}
\label{secResDerElMod}

\subsubsection{Activation energy at the CB peak}
\label{secActEnCBPeak}
With lowering the temperature, the SET conductance due to sequential tunneling
is suppressed as
\begin{equation}
G\propto \exp\left(-\Delta/T\right),
\label{supplGDT}
\end{equation}
where $\Delta$ is an activation energy and $T$ is the temperature.
The activation energy is the excitation energy of the 
SET required in order to transfer an electron from the SET island into the lead or
vice versa.
Away from the anomaly region, the activation energy, around a sequential-tunneling peak, 
is given by
\begin{equation}
\Delta = \left|E(n,N)-E(n,N+1)\right|,
\label{supplDelta1}
\end{equation} 
because the transport occurs due to the fluctuation $N\leftrightarrow N+1$ at fixed $n=0$ (at $N_g<N_g^-$) or $n=1$ (at $N_g>N_g^+$).
According to Eq.~(\ref{supplDelta1}), the activation energy 
changes linearly, on both sides of the peak, 
from the smallest value, $\Delta=0$, at the peak center,
to the largest value, $\Delta=E_C$, in the middle of each of the two adjacent CB valleys.
However, as one approaches the anomaly region,
the lower and the upper bounds of $\Delta$ change.

\begin{figure}[th]
\begin{center}
\includegraphics[width=0.47\columnwidth]{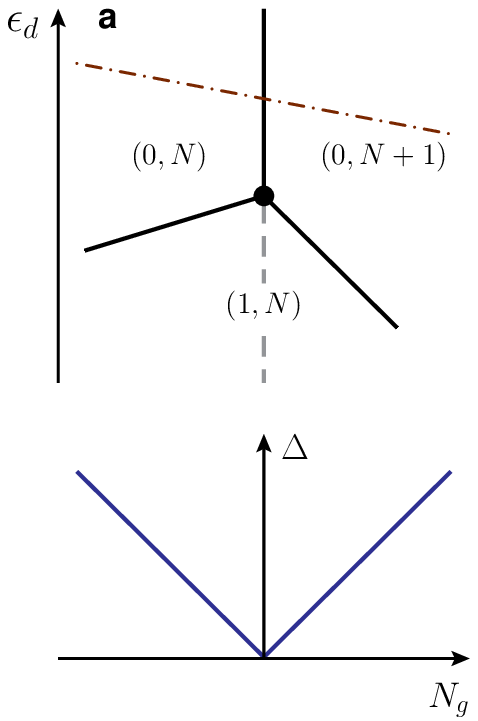}\hspace{12pt}
\includegraphics[width=0.47\columnwidth]{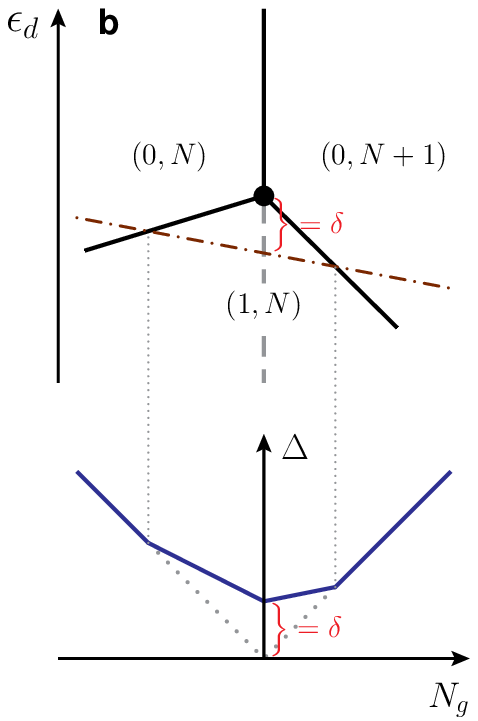}
\caption{
Activation energy $\Delta$ around a sequential-tunneling peak.
(a)
Upper panel: fragment of the stability diagram, see Fig.~\ref{fig2}b,
showing the working line (dash-dotted line) outside the anomaly region.
Lower panel: the $\Delta$ {\it vs.} $N_g$ dependence, corresponding to the upper panel.
(b)
Same as in (a), but with the working line in the anomaly region.
The smallest value of $\Delta$ is $\delta$.
}
\label{figS1}
\end{center}
\end{figure}

We illustrate how $\Delta$ is modified around the lower bound in Fig.~\ref{figS1}.
The left panel shows the usual behavior away from the
anomaly region.
The right panel shows the modification of $\Delta$ and 
the occurrence of a different lower bound, $\delta$,
in the anomaly region. 
Indeed, since the ground state is $(1,N)$ and the closest
transport channel is $(0,N)\leftrightarrow (0,N+1)$,
then transport is possible only if the donor 
happens to be in an excited state, corresponding to charge $n=0$, 
instead of the lowest energy state ($n=1$).
Thus, to the left of the peak center, we have
\begin{equation}
\Delta = E(0,N+1) - E(1,N),
\label{supplD1}
\end{equation}
whereas, to the right, we have 
\begin{equation}
\Delta = E(0,N) - E(1,N).
\label{supplD2}
\end{equation}
This modification of the activation energy, as compared with Eq.~(\ref{supplDelta1}), 
results in strongly asymmetric CB peaks at low temperatures.
The peaks become sharper on the side which extends away from the center of the anomaly.
The peaks maxima are still given by $E(n,N)=E(n,N+1)$,
as long as the temperature is sufficiently low ($T\ll U_{12}^2/E_C$).
The conductance at the peak maximum is suppressed as
\begin{equation}
G\propto \exp\left(-\delta/T\right),
\end{equation}
where $\delta$ is the lower bound of the activation energy $\Delta$.
From Eqs.~(\ref{supplD1}) and (\ref{supplD2}), at the peak maximum,  we obtain
\begin{equation}
\delta = \left.\left[E(0,N) - E(1,N)\right]\right|_{E(0,N)=E(0,N+1)}.
\label{suppld1}
\end{equation}
Equation~(\ref{suppld1}) is valid on the left-hand side of the anomaly,
$N_g^-\leq N_g\leq (N_g^++N_g^-)/2$.
On the right-hand side, a similar derivation holds, with respect to the
lower triple point, and $\delta$ is given by
\begin{equation}
\delta = \left.\left[E(1,N+1) - E(0,N+1)\right]\right|_{E(1,N)=E(1,N+1)},
\label{suppld2}
\end{equation}
for $(N_g^++N_g^-)/2\leq N_g\leq N_g^+$.
In the middle of the anomaly, $\delta$ takes on the largest value, given by
\begin{equation}
\delta_{\rm max} = \frac{U_{12}}{2}\left(1-\frac{U_{12}}{2E_C}\right).
\label{suppld3}
\end{equation}
Here and throughout the paper we make the reasonable assumption that $U_{12}<2E_C$.

For simplicity, let us assume that $N_g$ is measured with respect to the middle of the anomaly, i.e. $N_g^-=-N_g^+$.
Then, an explicit expression for $\delta$ reads
\begin{equation}
\delta(N_g) = \delta_{\rm max}\left(1 -\frac{\left|N_g\right|}{N_g^+}\right)\theta(N_g^+-\left|N_g\right|),
\label{suppld4}
\end{equation}
where $\theta(x)=1$ for $x>0$ and $\theta(x)=0$ for $x<0$.
Equation~(\ref{suppld4}) is shown in Fig.~\ref{fig2}c as the solid line.

In the special case $g_L=0$, the behavior of $\delta$ is modified outside the anomaly region.
Indeed, transport is possible, in this case, only due to hopping over the donor.
Activationless transport can occur only at the triple points,
where two resonance conditions are met,
\begin{equation}
\left\{
\begin{array}{ccc}
E(n,N)&=&E(n,N+1),\\
E(1,N)&=&E(0,N+1),
\end{array}
\right.
\end{equation}
both for the upper ($n=0$) and lower ($n=1$) triple points.
In this case, we find that Eq.~(\ref{suppld4}) for $\delta$ changes as follows
\begin{equation}
\delta(N_g) = \delta_{\rm max}\left(1 -\frac{\left|N_g\right|}{N_g^+}\right)\left[2\theta(N_g^+-\left|N_g\right|)-1\right].
\label{suppld5}
\end{equation}
Equation~(\ref{suppld5}) is shown in Fig.~\ref{fig2}c as the dashed line.

\subsubsection{Activation energy in the CB valley}
\label{secActEnCBValley}
In Sec.~\ref{secActEnCBPeak}, we discussed the lower bound of the activation energy $\Delta$, see Eq.~(\ref{supplGDT}).
Here we focus on the upper bound of $\Delta$ and discuss the conductance in the CB valleys.
We assume that the tunnel coupling is sufficiently small and, therefore, 
we discard cotunneling.

We start with a position $N_g$,
deep in the region of usual behaviour (for concreteness that with $n=0$),
where the activation energy is given by Eq.~(\ref{supplDelta1})
and assumes $\Delta=E_C$ in the middle of a CB valley, say at $N_g=N$.
It is important to remark that while the position $N_g=N$ corresponds to
the middle of a CB valley for the $n=0$ branch of resonances, 
this position is off the middle for the $n=1$ branch. 
Thus, from the prospect of the $n=1$ branch, the position $N_g=N$ has a smaller
activation energy because of the branch shift
$N_g\to N_g+U_{12}/2E_C$.
Therefore, if $n$ would change from $0$ to $1$, a transport sequence with a smaller 
activation energy would be possible. 
However, changing $n$ from $0$ to $1$ also costs an activation energy.
The point at which the two factors balance each other is
\begin{equation}
\overline{N_g^+}=N_g^+\left(1+\frac{1}{1-U_{12}/2E_C}\right),
\end{equation}
where we assume, as before, that $N_g$ is measured with respect to the center of the anomaly.
For $|N_g|<\overline{N_g^+}$, the energy spent to change the donor charge is less
than the gain in the activation energy of the transport sequence; and hence, $\Delta<E_C$.
The upper bound of $\Delta$ is given by
\begin{equation}
\overline{\delta}(N_g) = E_C-\frac{U_{12}}{2}
\left(1 -\frac{\left|N_g\right|}{\overline{N_g^+}}\right)\theta(\overline{N_g^+}-\left|N_g\right|).
\label{supplOOOd4}
\end{equation}
The positions of the minima shift linearly from $N_g=N$ to $N_g=N+U_{12}/2E_C$
as one traverses the upper-bound anomaly, $\overline{N_g^-}<N_g<\overline{N_g^+}$.

\subsubsection{Average occupation $\langle n\rangle$}
\label{supplSecnaver}
In Eq.~(\ref{linGnaver}),
we expressed the linear conductance $G(N_g)$ through the average occupation $\langle n\rangle$.
Here we give explicit expressions for $\langle n\rangle$ in different regimes.

First, let us find the average occupation $\langle n\rangle$ of an isolated donor 
at thermodynamic equilibrium with a lead at temperature $T$ and chemical potential $\mu=0$.
Following the general prescription, we have
\begin{eqnarray}
&&\langle n\rangle = 1\rho_\uparrow + 1 \rho_\downarrow + 0 \rho_0,\nonumber\\
&&Z=2 e^{-\epsilon_d/T} + 1,\nonumber\\
&&\rho_\uparrow=\rho_\downarrow= \frac{1}{Z}e^{-\epsilon_d/T},\nonumber\\
&&\rho_0=\frac{1}{Z}.
\end{eqnarray}
Thus, we obtain
\begin{equation}
\langle n\rangle=\frac{1}{1+(1/2)\exp(\epsilon_d/T)}.
\label{naverisol}
\end{equation}
In the expressions above, $\rho_0$, $\rho_\uparrow$, and $\rho_\downarrow$ 
denote, respectively, the probabilities for the donor to be in the empty, spin up, and spin down states;
$Z$ denotes the statistical sum.

Next, we find the average occupation $\langle n\rangle$ of the donor in the presence of the SET, 
as used in Eq.~(\ref{linGnaver}).
The general expression reads
\begin{eqnarray}
&&\langle n\rangle = \sum_{nN}n\rho_{nN},
\label{navergeneral}
\\
\label{supplrhonNgeneral}
&&\rho_{nN}=\frac{1}{Z}s_ne^{-E(n,N)/T},
\end{eqnarray}
where $Z=\sum_{nN}s_ne^{-E(n,N)/T}$ is the statistical sum,
$s_n$ is the spin degeneracy on the donor ($s_0=1$ and $s_1=2$),
and $E(n,N)$ is given in Eq.~(\ref{eqelectrostatic}).
Equation~(\ref{navergeneral}) can be given a form similar to Eq.~(\ref{naverisol}),
\begin{equation}
\langle n\rangle=\frac{1}{1+(Z_-/2Z_+)\exp\left[(\epsilon_d-\epsilon_d^0)/T\right]},
\label{naver0}
\end{equation}
where $\epsilon_d^0=\frac{U_{12}^2}{4E_C}$ is the middle position in the anomaly,
$Z_-=\sum_Ne^{-E_C(N-N_g)^2/T}$ is the statistical sum of the SET on the left side of the anomaly,
and, correspondingly, $Z_+=\sum_Ne^{-E_C(N-N_g+U_{12}/2E_C)^2/T}$
is the statistical sum of the SET on the right side of the anomaly.
We remark that, in the limit $U_{12}\ll T$,
Eq.~(\ref{naver0}) reduces to Eq.~(\ref{naverisol}).

Next, we consider the vicinity of a pair of triple points,
as shown in the inset of Fig.~\ref{fig2}b. 
We assume $T, U_{12}\ll E_C$.
The relevant charge states of the SET are $N$ and $N+1$.
Then, the factor $Z_-/Z_+$ in Eq.~(\ref{naver0}) reduces to
\begin{eqnarray}
\frac{Z_-}{Z_+}&=&
\frac{
\cosh\left[\frac{E_C}{T}\left(N-N_g+\frac{1}{2}\right)\right]}{
\cosh\left[\frac{E_C}{T}\left(N-N_g+\frac{1}{2}+\frac{U_{12}}{2E_C}\right)\right]}\nonumber\\
&&\times
e^{\frac{U_{12}}{T}\left(N-N_g+\frac{1}{2}+\frac{U_{12}}{4E_C}\right)}.
\label{naver1}
\end{eqnarray}
Equation~(\ref{naver1}) together with Eq.~(\ref{naver0}) give the average occupation $\langle n\rangle$
in the vicinity of a pair of triple points.

Around the upper triple point the charge configuration $(1,N+1)$ can be omitted and
the expression for $\langle n\rangle$ can be simplified further,
\begin{equation}
\langle n\rangle = \left[1+\frac{1}{2}\left(1+e^{2E_C\delta N_g/T}\right)e^{(\delta\epsilon_d-U_{12}\delta N_g)/T}\right]^{-1},
\label{supplavn1UTP}
\end{equation}
where $\delta N_g=N_g -N-1/2$ and $\delta\epsilon_d = \epsilon_d-U_{12}/2$ are, respectively, 
the dimensionless gate voltage and the donor energy
measured from the upper triple point.
Similarly, around the lower triple point the average occupation $\langle n\rangle$ can
be written as follows
\begin{equation}
\langle n\rangle = \left[1+\frac{1}{2}\left(1+e^{2E_C\overline{\delta N}_g/T}\right)^{-1}
e^{(\overline{\delta\epsilon}_d-U_{12}\overline{\delta N}_g)/T}\right]^{-1},
\label{supplavn1LTP}
\end{equation}
where $\overline{\delta N}_g=N_g -N-1/2-U_{12}/2E_C$ and $\overline{\delta\epsilon}_d = \epsilon_d+U_{12}/2-U_{12}^2/2E_C$ are, respectively, 
the dimensionless gate voltage and the donor energy
measured from the lower triple point.

Note that at the upper triple point the average charge is $\langle n\rangle=1/2$, whereas at the lower triple point it is
$\langle n\rangle=4/5$.
This asymmetry is due to the spin degeneracy on the donor. 

\subsection{Kinetic theory of conduction}
\label{secTransport}
In Sec.~\ref{secResDerElMod}, we discussed the activational conductance through the donor-SET system
and related the activation energy $\Delta$ to the electrostatic model of Eq.~(\ref{eqelectrostatic}).
Such an analysis helps to understand the qualitative behaviour of the conductance $G(V_g)$
in the anomaly region.
A more rigorous consideration of $G(V_g)$, would involve inclusion of a prefactor
in front of the exponential dependence in Eq.~(\ref{supplGDT}).
For large $\Delta/T$ (i.e. for small temperatures), 
such a prefactor is of little relevance, since most features in $G(V_g)$ come from 
the dependence of $\Delta$ on $V_g$, and are enhanced by the fact that $\Delta$ appears in the exponent.
At higher temperatures, the dependence of the prefactor on $V_g$ can be comparable 
to the dependence of $\exp\left(-\Delta/T\right)$ on $V_g$.
It turns out (see below) that the analysis made in Sec.~\ref{secResDerElMod} is valid at $T\ll U_{12}^2/E_C$.
In the rest of the paper, we give a more rigorous treatment of the problem.
We complete the electrostatic model of Eq.~(\ref{eqelectrostatic})
with kinetic-energy and tunneling terms, and give a rate-equations description of the transport problem.
We analyze different temperature regimes and find that, at $T\ll U_{12}$,
the rate-equations description
is identical to a description in which the actual (interacting) system is replaced by 
a circuit of resistors.
We employ such a replacement as an approximation at any temperature
and illustrate its accuracy for the parameters relevant to the experiment.

\subsubsection{Generalities on the Hamiltonian and sequential tunneling theory}
\label{secGenHamSeqTunThr}
We assume that transport through the SET is due to sequential tunneling.~\cite{GlazmanShekhter,Beenakker,AverinKorotkovLikharev}
In order to calculate the linear conductance, we rewrite
the model in Eq.~(\ref{eqelectrostatic})
in second-quantized form and complement it with
tunneling terms.
The donor Hamiltonian takes the form
\begin{equation}
H_d = \epsilon_d \sum_{\sigma=\uparrow,\downarrow} d_\sigma^\dagger d_\sigma +
U_\infty n_\uparrow n_\downarrow,
\end{equation}
where $d_\sigma^\dagger$ ($d_\sigma$) is the creation (annihilation) operator of an electron of spin $\sigma=\{\uparrow,\downarrow\}$ on the donor.
The donor occupation is given by
$n=n_\uparrow + n_\downarrow$, where
$n_\sigma=d_\sigma^\dagger d_\sigma$ is the occupation of one spin species.
The on-site repulsion energy $U_\infty$ on the donor is assumed to be large enough in order to exclude the occupation $n=2$.
The SET island is described by
\begin{equation}
H_D = \sum_{k\sigma}\varepsilon_k f_{k\sigma}^\dagger f_{k\sigma} + E_C\left(N-N_g\right)^2,
\label{eqHDSETgen}
\end{equation}
where $\varepsilon_k$ is the energy spectrum in the island,
$f_{k\sigma}^\dagger$ ($f_{k\sigma}$) is the creation (annihilation) operator of an electron in state 
$(k, \sigma)$ on the island, and
$N=\sum_{k\sigma}f_{k\sigma}^\dagger f_{k\sigma}$ is the operator of the number of electrons on the island.
The single-particle energy spacing in the island 
is assumed to be vanishingly small (much smaller than the temperature).
The electrostatic and tunnel coupling between donor and SET island is given by
\begin{equation}
H_{dD}= U_{12}\left(N-N_g\right)\sum_{\sigma}n_\sigma  +
t_{12}\sum_{k\sigma}\left(d_\sigma^\dagger f_{k\sigma}+{\rm h.c.}\right),
\label{eqHdD}
\end{equation}
where $t_{12}$ is the tunneling amplitude between donor and SET island.
The leads, i.e. the source ($L$) and the drain ($R$), are described by
\begin{equation}
H_{\rm leads}=\sum_{l=L,R}\sum_{p\sigma}\xi_p c_{lp\sigma}^\dagger c_{lp\sigma},
\end{equation}
where $\xi_p$ is the dispersion relation in the leads
and $c_{lp\sigma}^\dagger$ ($c_{lp\sigma}$) is the creation (annihilation) operator of an electron
with momentum $p$ and spin $\sigma$ in lead $l=L,R$.
The tunneling between the donor and the left lead is described by
\begin{equation}
H_{dL}=t_L\sum_{p\sigma}(d_{\sigma}^\dagger c_{Lp\sigma}+ {\rm h.c.}),
\label{eqHdL}
\end{equation}
where $t_L$ is the corresponding tunneling amplitude.
The tunneling between the SET island the leads is described by
\begin{equation}
H_{DLR}=\sum_{l}V_l\sum_{kp\sigma}(f_{k\sigma}^\dagger c_{lp\sigma}+ {\rm h.c.}),
\label{eqHDLR}
\end{equation}
where $V_l$ is the tunneling amplitude between SET island and lead $l=L,R$.

Next we calculate the sequential-tunneling rates for each of the tunneling terms in 
Eqs.~(\ref{eqHdD}), (\ref{eqHdL}), and (\ref{eqHDLR}).
The sequential tunneling between donor and SET is described by the rates
\begin{eqnarray}
W_{1N,0N+1}&=& \frac{4\Gamma_R}{\hbar}\nonumber\\
&\times&
f\left(\epsilon_d + U_{12}(N-N_g)
-E_C\left(2(N-N_g)+1\right)
\right),\nonumber\\
W_{0N+1,1N}&=& \frac{2\Gamma_R}{\hbar}
\left[1 \right.\nonumber\\
&-&\left. 
f\left(\epsilon_d + U_{12}(N-N_g)
-E_C\left(2(N-N_g)+1\right)
\right)\right],\nonumber\\
\label{supplW12}
\end{eqnarray}
where $\Gamma_R=\pi \nu_D \left|t_{12}\right|^2$, with $\nu_D$ being the density of states in the SET.
The different prefactors in the direct and reverse rates come from the spin degeneracy on the donor.
The sequential tunneling between donor and lead is described by the rates
\begin{eqnarray}
W_{1N,0N}&= &\frac{4\Gamma_L}{\hbar}
f\left(
\epsilon_d + U_{12}(N-N_g) - \Delta\mu_L
\right),\nonumber\\
W_{0N,1N}&=& \frac{2\Gamma_L}{\hbar}
\left[1-f\left(
\epsilon_d + U_{12}(N-N_g) - \Delta\mu_L
\right)\right],\nonumber\\
\label{W0N1N}
\end{eqnarray}
where $\Gamma_L=\pi \nu \left|t_L\right|^2$, with $\nu$ being the density of states in the lead,
and $\Delta\mu_l=\mu_l-\mu$, for $l=L,R$.
The sequential-tunneling rates of the SET at a fixed $n$ read
\begin{widetext}
\begin{eqnarray}
W_{nN+1,nN}&=& \sum_{l=L,R} W_{0N+1,0N}^l = \sum_{l}\frac{g_l}{\pi\hbar}
\Theta\left(\Delta\mu_l - 2E_C\left(N-N_g+1/2\right)-U_{12}n\right),\nonumber\\
W_{nN,nN+1}&=&\sum_{l=L,R} W_{0N,0N+1}^l = \sum_{l}\frac{g_l}{\pi\hbar}
\Theta\left(U_{12}n+2E_C\left(N-N_g+1/2\right)-\Delta\mu_l\right),
\label{supplWnnNN1}
\end{eqnarray}
where $g_l=4\pi^2\nu\nu_D\left|V_l\right|^2$ is the dimensionless conductance (in units of $e^2/\pi\hbar$) of lead $l$
and  $\Theta (E)=E/\left[1-\exp(-E/T)\right]$.

With the help of the rates in Eqs.~(\ref{supplW12}), (\ref{W0N1N}), and (\ref{supplWnnNN1}),
we are able to calculate the linear conductance of the donor-SET system in different regimes.

\subsubsection{Two triple points ($U_{12}, T \ll E_C$)}
\label{sec2trplPt}
We consider the vicinity of two triple points, as shown in the inset of Fig.~\ref{fig2}b,
and write down the rate equations for four charge configurations, $(0N,0N+1,1N,1N+1)$,
\begin{eqnarray}
\left(W_{0N+1,0N}+W_{1N,0N}\right)\rho_{0N}
&=&
W_{0N,0N+1}\rho_{0N+1}+W_{0N,1N}\rho_{1N}, \nonumber\\
\left(W_{0N,0N+1}+W_{1N,0N+1}+W_{1N+1,0N+1}\right)\rho_{0N+1}
&=&
W_{0N+1,0N}\rho_{0N}+W_{0N+1,1N}\rho_{1N}+W_{0N+1,1N+1}\rho_{1N+1}, \nonumber\\
\left(W_{0N,1N}+W_{0N+1,1N}+W_{1N+1,1N}\right)\rho_{1N}
&=&
W_{1N,0N}\rho_{0N}+W_{1N,0N+1}\rho_{0N+1}+W_{1N,1N+1}\rho_{1N+1}, \nonumber\\
\left(W_{0N+1,1N+1}+W_{1N,1N+1}\right)\rho_{1N+1}
&=&
W_{1N+1,0N+1}\rho_{0N+1}+W_{1N+1,1N}\rho_{1N}.
\label{fullMEqs}
\end{eqnarray}
\end{widetext}
These equations are complimented by the 
normalization condition $\rho_{0N}+\rho_{0N+1}+\rho_{1N}+\rho_{1N+1}=1$.
The current can be evaluated, for example, at the right tunnel junction,
\begin{eqnarray}
I&=&-|e|\left(
W^{R}_{0N,0N+1}\rho_{0N+1}+W^{R}_{1N,1N+1}\rho_{1N+1}\right.\nonumber\\
&&
\left.
-W^{R}_{0N+1,0N}\rho_{0N}-W^{R}_{1N+1,1N}\rho_{1N}\right).
\label{eqthecurrent}
\end{eqnarray}
The linear conductance is obtained as $G=-|e|dI/d\Delta\mu$ in the limit
$\Delta\mu\to 0$, where $\Delta\mu=\mu_L-\mu_R$.

We focus in the following on the case 
$\Gamma_R=0$ and assume, for simplicity, that $\Gamma_L\to 0$.
This is the case of interest with regard to Eq.~(\ref{linGnaver}).
Using the rates in Eq.~(\ref{supplWnnNN1}), we calculate 
the conductance through the SET for a fixed charge on the donor
and recover the well-known expression 
\begin{equation}
G(N_g)=\frac{e^2}{h}\frac{g_Lg_R}{g_L+g_R}\frac{x-x_N}{\sinh(x-x_N)},
\end{equation}
where $x=2E_CN_g/T$.
The position of the peak is given by $x_N=2E_C(N+1/2)/T$ for the donor in state
$n=0$ and by $x_N=\left[2E_C(N+1/2)+U_{12}\right]/T$ for the donor in state $n=1$.
In the limit $\Gamma_L\to 0$, the donor changes state seldom and the current
as a function of time represents a telegraph noise.
Such telegraph noise is observed in our samples~\cite{HofheinzEPJ}
for a range of the gate voltage,
consistent with the picture given here.

The probability for the donor to be in state $n=0$ is given by $p_0=1-\langle n \rangle$
and the probability to be in state $n=1$ is $p_1=\langle n \rangle$.
Here, $\langle n \rangle$ is the average occupation on the donor, considered in Sec.~\ref{supplSecnaver}.
The linear conductance is the average over the two realizations of the donor occupation,
\begin{equation}
G(N_g)=\frac{e^2}{h}\frac{g_Lg_R}{g_L+g_R}\sum_n
p_n\frac{x-x_{nN}}{\sinh(x-x_{nN})},
\label{supplGEq3av}
\end{equation}
where $x_{nN}=\left[2E_C(N+1/2)+U_{12}n\right]/T$.
We note that the only way how correlations between
the donor and the SET charges enter in Eq.~(\ref{supplGEq3av})
is through the average occupation $\langle n \rangle$, which contains
information about the feedback of the SET state on the donor occupation probability.
For $T\ll E_C$, we sum the right-hand side of Eq.~(\ref{supplGEq3av}) over $N$
and obtain Eq.~(\ref{linGnaver}).

The periodicity of CB oscillations is broken in the anomaly region.
The CB peaks occur at positions $N_g=N+1/2$ to the left of the anomaly and
at positions $N_g=N+1/2+U_{12}/2E_C$ to the right of the anomaly.
The question arises: How is the shift of CB peaks distributed in the anomaly region?
Let the CB peaks in the anomaly region occur at position
\begin{equation}
N_g=N+\frac{1}{2}+\varphi(N_g),
\end{equation}
where the phase shift $\varphi(N_g)$ changes from $0$ to $U_{12}/2E_C$ as one crosses over the anomaly.
We find that there exist three temperature regimes in which $\varphi(N_g)$ has different behaviour.

At low temperatures, $T\ll U_{12}^2/E_C$,
as we already mentioned in Sec.~I~A, 
the CB peaks in the anomaly region have asymmetric shape.
The conductance decays exponentially on both sides of the peak.
However, the exponents of this decay are not the same on both sides, see Fig.~\ref{figS1}.
The peak is sharper on the side that faces away from the center of the anomaly.
In the low-temperature limit ($T\to 0$), 
the position of the peak is determined by the smallest activation energy.
The dependence of $\varphi$ on $N_g$ is a step function, 
\begin{equation}
\varphi(N_g)=\frac{U_{12}}{2E_C}\theta(N_g),
\end{equation}
with the jump occurring in the middle of the anomaly.

At intermediate temperatures, $U_{12}^2/E_C\ll T\ll U_{12}$,
the activation energy $\Delta$, entering in Eq.~(\ref{supplGDT}),
can be considered to be constant in a region of $N_g$ where $\Delta$ varies weakly,
see Fig.~\ref{figS1}.
The shape of the CB peak is determined by the prefactor 
of the exponential dependence in Eq.~(\ref{supplGDT}) and
resembles the sharp angle of a right triangle: 
the conductance increases linearly, reaches maximum, and then drops abruptly.
The sharp side of the peak faces the center of the anomaly.
When we speak of the peak position, we understand the position
given by the first moment of $N_g$,
\begin{equation}
\langle N_g\rangle=\frac{\int N_gG(N_g)dN_g}{\int G(N_g)dN_g}.
\label{supplNgavdef}
\end{equation}
This position differs from $N+1/2$ by the shift
\begin{equation}
\varphi = \frac{U_{12}}{6E_C}
\left[
1+\frac{N_g}{N_g^+} +\theta(N_g)
\right]
\theta\left(N_g^+-\left|N_g\right|\right).
\label{supplvarphiregime21}
\end{equation}
Here we neglected small corrections proportional to temperature
in the vicinity of $N_g=N_g^\pm$ and a broadening of the step function
over a small region of $N_g$, given by $\left|N_g\right|\lesssim (T/U_{12})N_g^+$.
We remark that, if, instead of Eq.~(\ref{supplNgavdef}), we use the position of the
peak maximum in order to talk about the shift $\varphi$, then 
Eq.~(\ref{supplvarphiregime21}) is modified as follows,
\begin{equation}
\varphi = \frac{U_{12}}{4E_C}
\left(
1+\frac{N_g}{N_g^+}
\right)
\theta\left(N_g^+-\left|N_g\right|\right).
\label{supplvarphiregime22}
\end{equation}

At high temperatures, $T\gg U_{12}$,
the dependence of $\langle n\rangle$ on $N_g$ is weak.
Substituting Eq.~(\ref{supplGEq3av}) into Eq.~(\ref{supplNgavdef})
and taking $\langle n\rangle$ out of the sign of integration, we obtain
\begin{equation}
\varphi = \frac{U_{12}}{2E_C}\langle n\rangle.
\label{supplgateeffvarphi}
\end{equation}
An expression for $\langle n\rangle$ in this regime is given in Eq.~(\ref{naverisol}).
We remark that this result is consistent with the mean-field approach, in which
one takes the average of $n$ in the system Hamiltonian and then calculates the SET conductance.
The resulting effect is a shift of the conductance peaks by an amount given in 
Eq.~(\ref{supplgateeffvarphi}).
The shift of CB peaks is distributed over a region of $N_g$ 
which extends outside the anomaly.
In practice, such a shift can be detected only by comparing the shifts of neighbouring peaks.
For this purpose, the following expression is useful,
\begin{equation}
\frac{\partial \varphi}{\partial N_g}=\frac{\alpha U_{12}}{2T}\langle n\rangle\left(1-\langle n\rangle\right).
\label{supplpartvarphipartN_g}
\end{equation}
Furthermore, the sample may also contain donors which are situated further away from the SET, and therefore, also from the top gate.
For such donors, the quantity in Eq.~(\ref{supplpartvarphipartN_g}) can be neglected, since 
both $\alpha$ and $U_{12}$ are small.
The leading-order effect comes from Eq.~(\ref{supplgateeffvarphi}) 
and represents a shift $\varphi_0$ 
which is constant as a function of $N_g$.
We expect $\varphi_0$ to have activational dependence on temperature,
$\varphi_0\propto \exp\left(-\Delta_{\rm imp}/T\right)$, with
$\Delta_{\rm imp}$ being the activation energy of the distant donors.

\begin{figure}[t!]
\begin{center}
\includegraphics[width=\columnwidth]{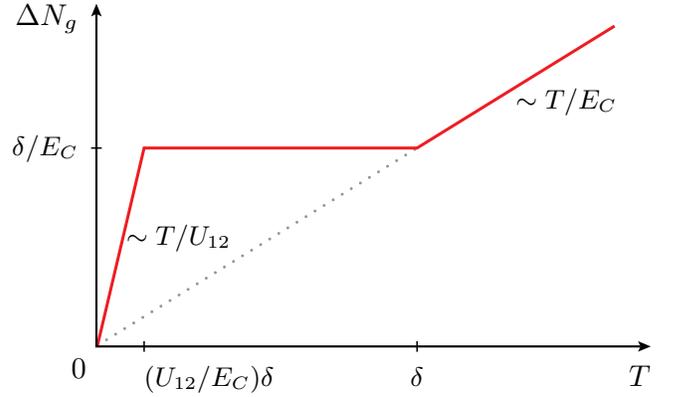}
\caption{
The width of the CB peak in the anomaly region
as a function of temperature.
}
\label{figS2}
\end{center}
\end{figure}

Next, we discuss the width of the CB peak as a function of temperature.
Outside the anomaly region the width is proportional to the temperature, 
$\Delta N_g\sim T/E_C$.
At high temperatures, $T\gg U_{12}$, this results persists in the anomaly region.
At lower temperatures, however, the width of the CB peak is considerably larger in the anomaly region than it is outside.
We find that the width of the CB peak, depends on the position $N_g$ in the anomaly, $N_g^- < N_g < N_g^+$.
This dependence enters through the dependence of $\delta$ on $N_g$, see Sec.~I~A.
We summarize our results in Fig.~\ref{figS2}.
At intermediate temperatures, $(U_{12}/E_C)\delta \ll T\ll\delta$,
the width is temperature-independent and given by $\Delta N_g\sim \delta/E_C$. 
At low temperatures, $T\ll (U_{12}/E_C)\delta$, the width is again proportional to temperature, 
but with a different proportionality coefficient, $\Delta N_g\sim T/U_{12}$.

\subsubsection{One triple point ($T\ll U_{12},E_C$)}
\label{SupplSec1TrPt}
We now focus on the upper triple point, retaining three charge states, 
$(0N,0N+1,1N)$, see Fig.~\ref{figS1}.
Around a single triple point, 
Eqs.~(\ref{fullMEqs}) and (\ref{eqthecurrent})
allow a more explicit consideration.
Quite remarkably, the end result of this lengthy calculation
is an expression which can also be obtained
by assigning an effective resistance to each tunnel junction
and replacing the studied system by a circuit of resistors, 
which is in the spirit of Ref.~\onlinecite{Ruzin}.
\begin{figure}[th]
\begin{center}
\includegraphics[width=\columnwidth]{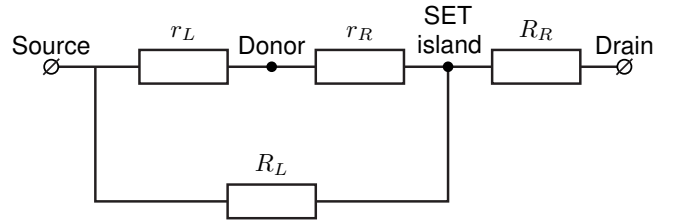}
\caption{
Circuit of resistors, representing the donor-SET system at low temperatures.
}
\label{figS3}
\end{center}
\end{figure}

The donor-SET system can, thus, be represented by the circuit in Fig.~\ref{figS3}.
The resistances associated with each of the tunnelling terms  
(see rates in Eqs.~(\ref{supplW12}), (\ref{W0N1N}), and (\ref{supplWnnNN1}))
are given by
\begin{eqnarray}
\frac{1}{r_R}&=&\frac{e^2}{T}W_{1N,0N+1}\rho_{0N+1}=\frac{e^2}{T}W_{0N+1,1N}\rho_{1N},\nonumber\\
\frac{1}{r_L}&=&\frac{e^2}{T}W_{1N,0N}\rho_{0N}=\frac{e^2}{T}W_{0N,1N}\rho_{1N},\nonumber\\
\frac{1}{R_l}&=&\frac{e^2}{T}W_{0N+1,0N}^l\rho_{0N}=\frac{e^2}{T}W_{0N,0N+1}^l\rho_{0N+1},\nonumber\\
\label{defRRRLR12Wrho}
\end{eqnarray}
where all the quantities are evaluated at thermodynamic equilibrium.
We note that each of the expressions in Eq.~(\ref{defRRRLR12Wrho}) has the form of the Einstein relation,
\begin{equation}
G=\frac{e^2D}{T},
\end{equation}
where the role of the diffusion constant $D$ is played by the product $W\rho$ with appropriate indices.
The total conductance follows from adding the resistances in Fig.~\ref{figS3},
\begin{equation}
\frac{1}{G}=R_R+\frac{R_L\left(r_L+r_R\right)}{R_L+r_L+r_R}.
\label{suppl1oGR}
\end{equation}
To give explicit expressions for the resistances in Eq.~(\ref{defRRRLR12Wrho}), we measure energy with respect to the
charge configuration $(0,N)$ and obtain
\begin{eqnarray}
\rho_{0N}&=&\frac{1}{Z},
\nonumber\\
\rho_{0N+1}&=&\frac{1}{Z}e^{2E_C\delta N_g/T},
\nonumber\\
\rho_{1N}&=&\frac{1}{Z}2e^{-(\delta\epsilon_d-U_{12}\delta N_g)/T},
\label{eqrho1tp2eq}
\end{eqnarray}
where $Z$ is the statistical sum given by
\begin{equation}
Z=1+2e^{-(\delta\epsilon_d-U_{12}\delta N_g)/T}+e^{2E_C\delta N_g/T}.
\label{supplZexpl}
\end{equation}
For the notations of 
$\delta\epsilon$ and $\delta N_g$, 
see expressions in the text below Eq.~(\ref{supplavn1UTP}).
The final expressions for resistances read
\begin{eqnarray}
r_R&=&\frac{\hbar}{e^2}\frac{Z T}{4\Gamma_{R}}\left(e^{(\delta\epsilon_d-U_{12}\delta N_g)/T} +e^{-2E_C\delta N_g/T}\right),\nonumber\\
r_L&=&\frac{\hbar}{e^2}\frac{Z T}{4\Gamma_L}\left(e^{(\delta\epsilon_d-U_{12}\delta N_g)/T}+1\right)
,\nonumber\\
R_l&=&\frac{\hbar}{e^2}\frac{\pi ZT}{g_l}
\frac{1-e^{-2E_C\delta N_g/T}}{2E_C\delta N_g}.
\label{defRRRLR12}
\end{eqnarray}

In the rest of this section, we derive Eq.~(\ref{eqrestunY}). 
We set $g_L=0$ and assume $g_R\gg \Gamma_l/T$.
In this case,
the linear conductance is determined by 
the resistance of the donor tunnel junctions,
$1/G=r_L+r_R$, see Eq.~(\ref{suppl1oGR}).
The linear conductance is largest at the triple points 
(or close to the triple points within a distance $\propto T$).
The CB oscillations show, therefore, a resonance
in the envelope function of the peaks
when the working line of the device intersects 
either the upper or the lower row of triple points.
The questions of interest, here, are: What is the shape of the envelope 
function and how does the maximal value of conductance at the resonance 
scale with temperature?

The envelope function can be determined by, first, finding the
positions of the maxima of the CB peaks, and then, evaluating the conductance
at these points.
With the simplifying assumptions we made above, 
the positions of the maxima are determined by solving
\begin{equation}
\frac{\partial}{\partial\, \delta N_g}\left(r_L+r_R\right)=0
\label{supplpartpartRR0}
\end{equation}
with respect to $\delta N_g$.
This task, however, makes the end result cumbersome and difficult to discuss.
We aim, instead, at finding a representative curve for the envelope function,
i.e. a curve which is correct by order of magnitude.
We find that, for $\Gamma_R\lesssim \Gamma_L$, the solution of Eq.~(\ref{supplpartpartRR0})
can be approximated as $\delta N_g\approx 0$.
Setting $\delta N_g=0$ in the expressions for $r_L$ and $r_R$ in Eq.~(\ref{defRRRLR12})
and using Eq.~(\ref{supplZexpl}), we obtain
\begin{equation}
\frac{1}{G}=\frac{\hbar T}{2e^2}\frac{\Gamma_L+\Gamma_R}{\Gamma_L\Gamma_R}
\left(1+e^{-\delta\epsilon_d/T}\right)
\left(e^{\delta\epsilon_d/T}+1\right),
\label{eqsuppl1oGenv0}
\end{equation}
which derives Eq.~(\ref{eqrestunY}), provided we also
relate $\epsilon_d$ to $N_g$ using the ``working line'' relation.

Focusing on the lower triple point, i.e. 
retaining only the charge states $(0N+1,1N,1N+1)$, 
we derive similar expressions to those above.
In particular, Eq.~(\ref{eqsuppl1oGenv0}) becomes
\begin{equation}
\frac{1}{G}=\frac{\hbar T}{4e^2}\frac{\Gamma_L+\Gamma_R}{\Gamma_L\Gamma_R}
\left(4+e^{\overline{\delta\epsilon_d}/T}\right)
\left(e^{-\overline{\delta\epsilon_d}/T}+1\right),
\label{eqsuppl1oGenv1}
\end{equation}
where $\overline{\delta\epsilon_d}$ is defined in the text below Eq.~(\ref{supplavn1LTP}).

We conclude this section by remarking that 
(i) the envelope function of the CB oscillations resembles the resonance transmission of the donor 
and 
(ii) the temperature dependence of the maximum of the envelope function is $G\propto 1/T$.

\subsubsection{Discussion of a ``sum-of-resistance'' formula for arbitrary case}
\label{secSumOfResFrom}
Here, we derive an approximate expression for the linear conductance by extending
the circuit approach of Sec.~\ref{SupplSec1TrPt} to arbitrary temperatures.
We remark that the circuit approach is a quick and computationally-inexpensive
way to obtain an insight into the behaviour of the linear conductance.
The approximation made when replacing the system under consideration by a circuit of
resistors consists in neglecting non-local correlations, i.e. correlations extending 
over several tunnel junctions.
We compare the circuit approach against a numerically exact evaluation of the conductance
at the end of this section.
Below, we derive expressions for the resistances of the resistors in Fig.~\ref{figS3}.

We consider a weak deviation from equilibrium, for which it is not necessary to
correct the distribution function.
The current flowing from the lead $l=L,R$ into the SET island
originates from the fluctuation $(n,N)\leftrightarrow (n,N+1)$
and can be written as
\begin{widetext}
\begin{eqnarray}
\label{eq:1}
I_l&=&-\frac{eg_l}{\pi\hbar}\sum_{nN}
\int d\varepsilon
\int d\varepsilon'
\left\{
\rho_{nN} f(\varepsilon)[1-f(\varepsilon')]
-
\rho_{nN+1} [1-f(\varepsilon)]f(\varepsilon')
\right\}
\nonumber\\
&&
\times
\delta(\varepsilon-\varepsilon'+\Delta\mu_l+E(n,N)-E(n,N+1)),
\label{suppleqIlcurr0}
\end{eqnarray}
\end{widetext}
where
$\rho_{nN}$ is the equilibrium distribution function, given in Eq.~(\ref{supplrhonNgeneral}),
and $\Delta\mu_l\to 0$.
Then, the resistance $R_l$ associated with junction $l$ is found from Eq.~(\ref{suppleqIlcurr0}) to be
\begin{equation}
\label{eq:2}
\frac{1}{R_l}
=
\frac{e^2}{\pi\hbar}
\frac{g_l}{T}
\sum_{nN}
 \rho_{nN}\Theta(E(n,N)-E(n,N+1)).
\end{equation}
In the same fashion, we write down the current
flowing from the SET island to the donor,
considering the fluctuation
$(0,N+1)\leftrightarrow(1,N)$,
\begin{eqnarray}
\label{eq:4}
i_R&=&-\frac{e\Gamma_R}{\hbar}
\sum_{N}
\int d\varepsilon
\left\{
 \rho_{1N} [1-f(\varepsilon)]
-2 \rho_{0N+1} f(\varepsilon)
\right\}
\nonumber\\
&&\times
\delta\left(
\varepsilon+\Delta\mu_D+E(0,N+1)-E(1,N)
\right).
\end{eqnarray}
Here, the factor of $2$ in front of $\rho_{0N+1}$  originates from the spin degeneracy on the donor 
and we assumed that the SET island is at an elevated chemical potential
$\Delta\mu_D\to 0$.
The resistance $r_R$ associated with this junction is, then, found to be
\begin{equation}
\label{eq:5}
\frac{1}{r_R}
=
\frac{e^2}{\hbar}
\frac{4\Gamma_R}{T}
\sum_{N}
 \rho_{0N+1}f \left( E(1,N)-E(0,N+1) \right) .
\end{equation}
Similarly, the resistance $r_L$ associated with the junction between the donor and the left lead is obtained from
\begin{equation}
\label{eq:6}
\frac{1}{r_L}
=
\frac{e^2}{\hbar}
\frac{4\Gamma_L}{T}
\sum_{N}
 \rho_{0N}f \left( E(1,N)-E(0,N) \right).
\end{equation}
With the help of Eqs.~(\ref{eq:2}), (\ref{eq:5}), and (\ref{eq:6}),
the conductance $G$ is given by Eq.~(\ref{suppl1oGR}).

\begin{table*}[th]
\begin{tabular*}{0.5\textwidth}{@{\extracolsep{\fill}} l c l l c l l}
\toprule[0.9pt]
\addlinespace[2pt]
&& \multicolumn{2}{c}{Sample 1}& & \multicolumn{2}{c}{Sample 2}\\
\addlinespace[1pt]
\cmidrule[0.6pt](r){3-4}
\cmidrule[0.6pt](r){6-7}
\addlinespace[2pt]
$T$   && $1.0\,{\rm K}$ &  $4.2\,{\rm K}$& & $0.9\,{\rm K}$ & $4.2\,{\rm K}$ \\
\addlinespace[2pt]
$E_C$ &&  $1.1\,{\rm meV}$ & $0.95\,{\rm meV}$ && $0.9\,{\rm meV}$ & $1.15\,{\rm meV}$ \\
\addlinespace[2pt]
$g_L$ && $0.017$ & $0.021$ && $0.01$ & $0.011$ \\
\addlinespace[2pt]
$g_R$ && $0.017$ & $0.021$ && $0.075$ & $0.095$\\
\addlinespace[2pt]
$U_{12}$ && $1.2\,{\rm meV}$ & $1.2\,{\rm meV}$  && $1.15\,{\rm meV}$ & $1.15\,{\rm meV}$ \\
\addlinespace[2pt]
$\Gamma_L$ && $0$ & $0$ && $40\,\mu{\rm eV}$ & $40\,\mu{\rm eV}$ \\
\addlinespace[2pt]
$\Gamma_R$ && $0$ & $0$ && $40\,\mu{\rm eV}$ & $40\,\mu{\rm eV}$\\
\addlinespace[2pt]
$\alpha$ && $0.077$ & $0.077$ && $0.165$ & $0.165$\\
\addlinespace[2pt]
$\beta$ && $0.45$ & $0.45$ && $0.55$ & $0.55$\\
\addlinespace[2pt]
$e/C_g$ && $0.01515\,{\rm V}$ & $0.01515\,{\rm V}$ && $0.011\,{\rm V}$ & $0.011\,{\rm V}$\\
\addlinespace[2pt]
$V_{g}(0)$ && $1.253\,{\rm V}$ & $1.264\,{\rm V}$ && $1.528\,{\rm V}$ & $1.534\,{\rm V}$\\
\addlinespace[1pt]
\bottomrule[0.9pt]
\end{tabular*}
\caption{
Parameters used in reproducing the experiment in Fig.~\ref{fig1}.
Samples 1 and 2 denote, respectively, the samples of Figs.~\ref{fig1}a and~\ref{fig1}b.
The parameters in each column are used in Eqs.~(\ref{suppl1oGR}), (\ref{eq:2}), (\ref{eq:5}), and (\ref{eq:6})  
to reproduce the measurement data at the corresponding temperature.
The result for Sample 1 (2) is shown in Fig.~\ref{fig1}c (Fig.~\ref{fig1}d).
Parameters $\alpha$ and $\beta$ determine the position and orientation of the working line
with the help of the relation
$\epsilon_d=U_{12}^2/4E_C - \alpha E_C (N_g - \beta)$.
The relation between $V_g$ and $N_g$ is conveniently parametrized as follows, $V_g=V_g(0)+(e/C_g)N_g$, where
the aim of the offset $V_g(0)$ is to restrict $N_g$ (and hence $N$) to small numbers.
}
\label{Table1}
\end{table*}

In Figs.~\ref{fig1}c and~\ref{fig1}d,
we reproduce the measurement data of Figs.~\ref{fig1}a and~\ref{fig1}b, respectively.
The parameters we have used
in Eqs.~(\ref{suppl1oGR}), (\ref{eq:2}), (\ref{eq:5}), and (\ref{eq:6})
are listed in Table~\ref{Table1}.

\begin{figure*}[th]
\begin{center}
\includegraphics[width=0.46\textwidth]{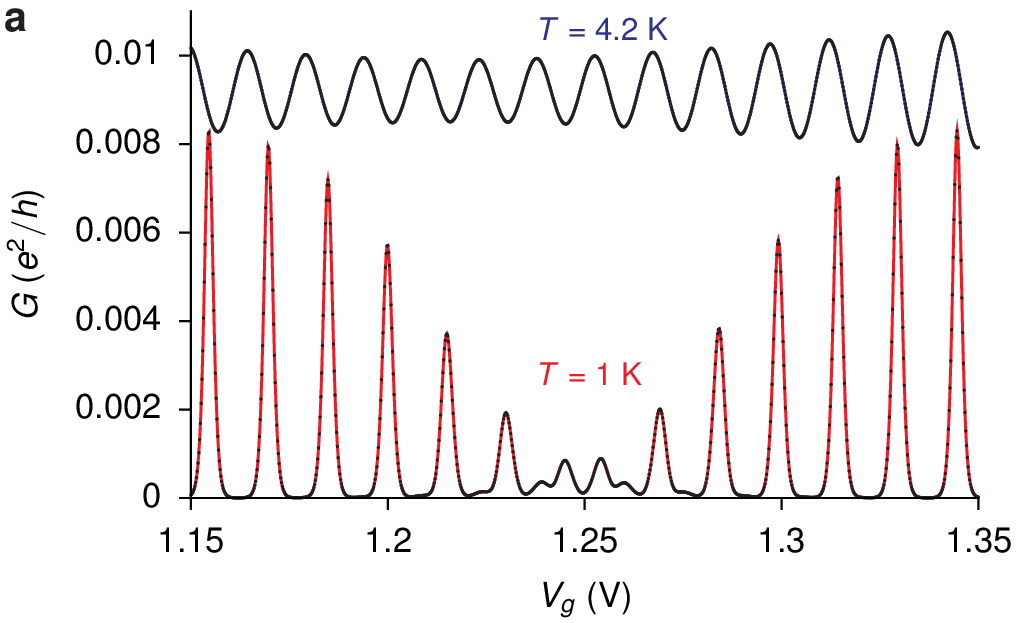}
\includegraphics[width=0.45\textwidth]{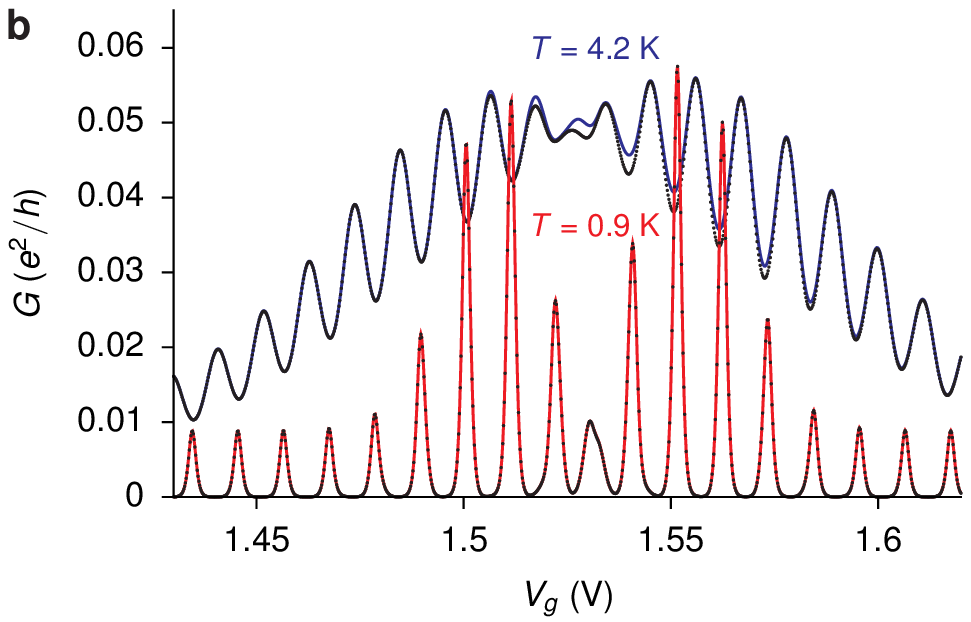}
\caption{
Comparison of the circuit approach against an exact numerical evaluation of the conductance.
The solid lines represent 
the result of the circuit approach, 
whereas the dots show the conductance
evaluated by solving the Pauli master equation numerically.
(a)
Conductance calculated for the same parameters as in Fig.~\ref{fig1}c.
The two approaches coincide identically in this case, because the circuit approach is exact
for the donor-SET model in the limit $\Gamma_l\to 0$, ($l=L,R$).
In particular, Eq.~(\ref{supplGEq3av}) can be obtained from Eqs.~(\ref{suppl1oGR}), (\ref{eq:2}), (\ref{eq:5}), and (\ref{eq:6}).
(b)
Conductance calculated for the same parameters as in Fig.~\ref{fig1}d.
For the low temperature trace ($T=0.9\;{\rm K}$), the circuit approach deviates from
the numerically exact result by a relative error $\varepsilon\leq 36\%$.
However, the deviation takes place far in the tails of the peaks where the conductance is small;
hence no visible difference on the scale of the plot.
For the high temperature trace ($T=4.2\;{\rm K}$), the
relative error is $\varepsilon\leq 7\%$ and the deviation is visible in several CB valleys. 
}
\label{figS4}
\end{center}
\end{figure*}

Next, we compare the circuit approach against an exact numerical evaluation of the conductance.
In Fig.~\ref{figS4}, we show Figs.~\ref{fig1}c and~\ref{fig1}d (solid line).
For the same set of parameters, we evaluate the conductance exactly (dots) 
by solving the Pauli master equation numerically.
In order to do this, 
we generalize the set of equations (\ref{fullMEqs}) to include all charge configurations $(n,N)$.
Then, we retain a sufficiently large number of charge configurations, around $N\approx N_g$,
to guarantee convergence, especially for $T\gtrsim E_C$.
We find good agreement between the two approaches for the parameters relevant to the experiment, see also caption of Fig.~\ref{figS4}.

\section{Conclusions}
\label{secConcl}
In conclusion, we have used a SET as a highly sensitive device to study single dopant atoms in Si.
A simple theoretical model allowed us to explain the anomalous behaviour of the linear conductance seen in the experiment
and to extract values of the coupling strength for the dopant atom and the SET.
Our results can be used to assess the feasibility of using a SET as a means
to manipulate and read out single dopant atoms in Si.

\begin{acknowledgments}
We thank Romain Wacquez and Maud Vinet at CEA/LETI for sample fabrication. 
We acknowledge financial support from  
the Nanosciences Fondation at Grenoble, France,
and the NSF grant DMR-0906498.
The research leading to these results has received funding
from the European Community's seventh Framework
FP7 2007/2013 under the Grant Agreement Nr: 214989.
\end{acknowledgments}

\appendix
\section{Stability diagram}
\label{appStabilityDiagram}
The stability diagram, see Fig.~\ref{fig2}b, shows regions in the two-parameter space $(N_g,\epsilon_d)$,
where the lowest-in-energy charge configurations of the donor-SET system is definite.
The solid lines separate the regions of different charge configurations, and hence, represent lines at which the
charge configuration is degenerate.
Using the electrostatic energy in Eq.~(\ref{eqelectrostatic}), we describe here
the stability diagram quantitatively.

The upper vertical lines in Fig.~\ref{fig2}b correspond to degeneracies of the type $E(0,N)=E(0,N+1)$.
The position of these degeneracies is given by
\begin{equation}
N_g=N+\frac{1}{2},
\end{equation}
where $N=0,\pm 1,\pm 2\dots$ takes on integer values.
The inclined lines with positive slope correspond to $E(0,N)=E(1,N)$ and are given by
\begin{equation}
\epsilon_d=U_{12}(N_g-N).
\end{equation}
The inclined lines with negative slope correspond to $E(0,N+1)=E(1,N)$ and are given by
\begin{equation}
\epsilon_d=E_C-(2E_C-U_{12})(N_g-N).
\end{equation}
The lower vertical lines correspond to $E(1,N)=E(1,N+1)$ and are given by
\begin{equation}
N_g=N+\frac{1}{2}+\frac{U_{12}}{2E_C}.
\end{equation}

The upper triple points correspond to the double condition $E(0,N)=E(1,N)=E(0,N+1)$ and have the coordinates
\begin{eqnarray}
N_g&=&N+\frac{1}{2},\nonumber\\
\epsilon_d&=&\frac{U_{12}}{2}.
\label{upperTPcoordinates}
\end{eqnarray}
Similarly, the lower triple points are found from $E(1,N+1)=E(1,N)=E(0,N+1)$ and have the coordinates
\begin{eqnarray}
N_g&=&N+\frac{1}{2}+\frac{U_{12}}{2E_C},\nonumber\\
\epsilon_d&=&-\frac{U_{12}}{2}+\frac{U_{12}^2}{2E_C}.
\label{lowerTPcoordinates}
\end{eqnarray}
From comparing Eqs.~(\ref{upperTPcoordinates}) and~(\ref{lowerTPcoordinates})
one finds that the middle position in the anomaly (middle between upper and lower triple points on $\epsilon_d$-axis)
is given by
\begin{equation}
\epsilon_d^0=\frac{U_{12}^2}{4E_C}.
\label{appAeps0mddl}
\end{equation}
Similarly, the width of the anomaly (distance between upper and lower triple points on $\epsilon_d$-axis)
is given by
\begin{equation}
\Delta\epsilon_d=U_{12} - \frac{U_{12}^2}{2E_C}.
\label{appADepsdWdth}
\end{equation}

Using the equation for the working line of the device, $\epsilon_d=-\alpha E_C N_g+{\rm const}$,
we relate $N_g^\pm$, see Fig.~\ref{fig2}b,
to the parameters of the stability diagram.
Considering the edges of the anomaly, we write
\begin{equation}
\epsilon_d^0 \mp \frac{\Delta\epsilon_d}{2} = -\alpha E_C N_g^\pm +{\rm const}.
\label{appAeqforNpmg}
\end{equation}
Together with Eqs.~(\ref{appAeps0mddl}) and~(\ref{appADepsdWdth}), the latter equation allows one to express $N_g^\pm$ 
in terms of the parameters $U_{12}$, $E_C$, $\alpha$, and the offset (${\rm const}$).
It is convenient for a practical purpose, to parametrize the offset by a parameter $\beta$, such that
the equation of the working line is written as follows
\begin{equation}
\epsilon_d=\frac{U_{12}^2}{4E_C} - \alpha E_C (N_g - \beta).
\end{equation}
The values of $\beta$ can be restricted to the interval $0\leq\beta<1$
by introducing an offset to the gate voltage $V_g$,
\begin{equation}
V_g=V_g(0)+\frac{eN_g}{C_g},
\end{equation}
where $V_g(0)$ is the gate voltage offset.
With this choice, the charge configuration with $N=0$ 
lies close to the center of the anomaly, and
the parameter $\beta$ gives the position of the anomaly center, $N_g=\beta$.
The anomaly edges are, then, given by
\begin{equation}
N_g^\pm=\beta \pm \frac{U_{12}}{2\alpha E_C}\left(1 - \frac{U_{12}}{2E_C}\right).
\label{appANgpmendres0}
\end{equation}
In the limit $\alpha\ll 1$, when discussing envelopes of CB oscillations, one may dispense with 
the term $\beta$ in Eq.~(\ref{appANgpmendres0}).

Finally, we remark that, for a fixed value of $\alpha$, the width of the anomaly is largest at $U_{12}=E_C$, see Eq.~(\ref{appANgpmendres0}),
and that the anomaly region shrinks to a point when $U_{12}$ approaches its largest value, $U_{12}=2E_C$.
The absence of the gapped region for $U_{12}=2E_C$ is related to the fact that the electrostatic model in 
Eq.~(\ref{eqelectrostatic}) can be rewritten as follows
\begin{equation}
E(n,N)=\left(\epsilon_d-E_C\right) n + E_C(n+N-N_g)^2,
\label{eqelectrostatic123}
\end{equation}
where we used the identity $n^2=n$.
The first term on the right-hand side in Eq.~(\ref{eqelectrostatic123}) 
can be interpreted as a kinetic energy term,
and thus, can be included into the kinetic-energy term of Eq.~(\ref{eqHDSETgen}) 
as an additional energy level in the SET island.
The second term represents the usual Coulomb energy of an island, with the difference that
the donor appears as part of that island too (the total charge being $n+N$).
It is important to note, however, that this analogy to the usual SET 
(and hence a seeming disappearance of the donor from the problem)
goes as far as the electrostatic energy is concerned and spin is ignored.
The donor may still leave visible traces in the transport, 
even at $U_{12}=2E_C$, provided both $\Gamma_L$ and $\Gamma_R$ are nonzero. 
As a matter-of-fact, the low-temperature trace of Fig.~\ref{figdata}a suggests precisely this scenario.


%

\end{document}